\documentclass[a4paper,11pt]{article}
\usepackage{siunitx}
\usepackage{jinstpub} 
\usepackage{mathtools}
\usepackage{setspace}
\usepackage{amsmath}
\usepackage{enumitem}
\newcounter{descriptcount}
\usepackage{hhline}
\usepackage{multirow}
\usepackage{textcomp}

\newlist{enumdescript}{description}{2}
\setlist[enumdescript,1]{%
  before={\setcounter{descriptcount}{0}%
          \renewcommand*\thedescriptcount{\alph{descriptcount}}}
  ,font=\bfseries\stepcounter{descriptcount}\thedescriptcount~
}
\setlist[enumdescript,2]{%
  before={\setcounter{descriptcount}{0}%
          \renewcommand*\thedescriptcount{\roman{descriptcount}}}
  ,font=\bfseries\stepcounter{descriptcount}\thedescriptcount~
}

\usepackage{lineno}
\usepackage{soul}

\newcommand{\reg}[1]{\textsuperscript{\small{\textregistered}} }

\newcommand{\internalref}[1]{}

\title{Development \& Characterization of Electrodes for large-scale Xenon Time Projection Chambers}

\abstract{
Dual-phase liquid xenon time projection chambers are the core detector elements of many experiments that conduct searches for Dark Matter and rare events, as well as in neutrino and high-energy physics. 
As part of this detector technology, high-voltage electrodes are instrumental for the generation of observable signals and their physical interpretation. 
Thus, electrode design and manufacturing has to fulfill stringent requirements, and their production is associated with significant engineering challenges. 
In this work we describe the successful development of electrodes on the \SI{1.5}{m}-scale, from their design and simulation to subsequent assembly and high-voltage testing in a gaseous argon environment. 
The produced electrodes were recently installed as an anode and a cathode during an upgrade to the XENONnT experiment.
}


\usepackage{xcolor}
\begin{document}


\author[a,1]{A.~Elykov,\note{Corresponding author.}}
\author[b,1]{S.~Vetter,}
\author[b,1]{V.~H.~S.~Wu,}
\author[c]{A.~Deisting,}
\author[b]{K.~Eitel,}
\author[b]{R.~Gumbsheimer,}
\author[b]{M.~Kara,}
\author[b]{S.~Lichter,}
\author[d]{S.~Lindemann,}
\author[d]{T.~Luce,}
\author[e]{Y.~Ma,}
\author[d]{J.~Müller,}
\author[b]{K.~Müller,}
\author[e]{K.~Ni,}
\author[c]{U.~Oberlack,}
\author[d]{M.~Schumann,}
\author[c]{P.~Shagin,}
\author[a,b]{K.~Valerius,}
\author[e]{M.~Zhong.}

\affiliation[a]{Institute of Experimental Particle  Physics, Karlsruhe Institute of Technology, 76021 Karlsruhe, Germany}
\affiliation[b]{Institute for Astroparticle Physics, Karlsruhe Institute of Technology, 76021 Karlsruhe, Germany}
\affiliation[c]{Institut für Physik \& Exzellenzcluster PRISMA+, Johannes Gutenberg-Universität Mainz, 55099 Mainz, Germany}
\affiliation[d]{Physikalisches Institut, Universität Freiburg, 79104 Freiburg, Germany}
\affiliation[e]{Department of Physics, University of California San Diego, La Jolla, CA 92093, USA}

\emailAdd{alexey.elykov@kit.edu}
\emailAdd{sebastian.vetter@kit.edu}
\emailAdd{hiu-sze.wu@kit.edu}

\maketitle
\flushbottom

\section{Introduction}
\label{intro}

Time projection chambers (TPCs) are commonly used detectors in particle and astroparticle physics. 
They range from extremely large to rather compact scales. 
Large-scale TPCs are exemplified by the DUNE experiment \cite{DUNE_TPC,instruments5040031} which plans to operate a modular liquid argon TPC of 70,000\,tonnes of total target mass for neutrino science and the ALICE \cite{ALME2010316} TPC with almost 90\,m$^3$ of a neon-based gas mixture for pattern recognition, tracking, and identification of charged particles in heavy-ion physics.
Other more compact realizations are high-pressure gaseous xenon TPCs such as NEXT \cite{next_epshep} with 100\,kg target mass and the AXEL \cite{HIKIDA2025170706} 180\,L-size prototype, both used for searches for neutrinoless double beta decay. 
In direct Dark Matter (DM) searches, TPCs are the central element of multiple experiments.
Among those is the DarkSide experiment, which featured a 50\,kg liquid argon target mass, and will be expanding it to 50\,tonnes \cite{DS-50, DS-20k}.
Furthermore, XENONnT \cite{nt_instrument}, LZ \cite{lz_detector} and PandaX-4T \cite{PandaX-4T-2019} are liquid-xenon (LXe) dual-phase TPCs with active masses of $\sim$\,6 and $\sim$\,7\,tonnes (XENONnT, LZ) and $\sim$\,4\,tonnes (PandaX-4T). 
Future TPCs in DM searches are envisaged to grow even larger \cite{Aalbers:2022dzr}, with PandaX-xT reaching 43\,tonnes with a TPC diameter of $\sim$\,2.6\,m \cite{PandaX-20T-2024}, and XLZD with 60 to 80\,tonnes of LXe target mass and a TPC measuring 3\,m in diameter \cite{XLZD:2024gxx}.

For the past decades, experiments based on LXe dual-phase TPC technology are among the leading searches in the field of DM detection, with one of their prime aims being the search for Weakly Interacting Massive Particles, which are seen as natural candidates for the role of DM \cite{Schumann:2019eaa}.
The unprecedentedly low background rates \cite{xenonnt_sr0_lower} coupled with low energy thresholds of $\sim$\,1\,keV allow such detectors to also search for a variety of other DM candidates \cite{xenonnt_se_dm,lz_ultraheavy}, detect astrophysical neutrinos \cite{xenonnt_b8}, perform searches for two-neutrino double electron capture (2$\nu$ECEC) and neutrinoless double beta decay (0$\nu\beta\beta$) \cite{xenon1t_2nuECEC, lz_0vbb_134xe}. 

Electrodes are a central element of LXe dual-phase TPCs and their design and high-voltage performance becomes an increasing challenge with ever larger TPC dimensions. 
In this paper, after a general introduction into the working principle and design of electrodes (\autoref{sec:signal_TPC} and \autoref{subsec:RnD_at_kit}) we present a novel and comprehensive approach to design, build and test large-scale electrodes.
We concentrate on two types of electrodes, a parallel-wire anode (\autoref{sec:parallel_wire}), and a  hexagonal mesh cathode (\autoref{sec:hex_mesh}), both with a diameter of $\sim$\,\SI{1.5}{m} typical for the XENONnT \cite{nt_instrument} and LZ \cite{lz_detector} experiments. 
In \autoref{sec:HV_test}, we report the high-voltage tests performed on the hexagonal mesh electrode as a method of quality assurance which can be generalized to other types of electrodes.

\subsection{Signals in Dual-phase Xenon Time Projection Chambers} 
\label{sec:signal_TPC}

\begin{figure}[htb]
\centering
\includegraphics[width=.8\textwidth]{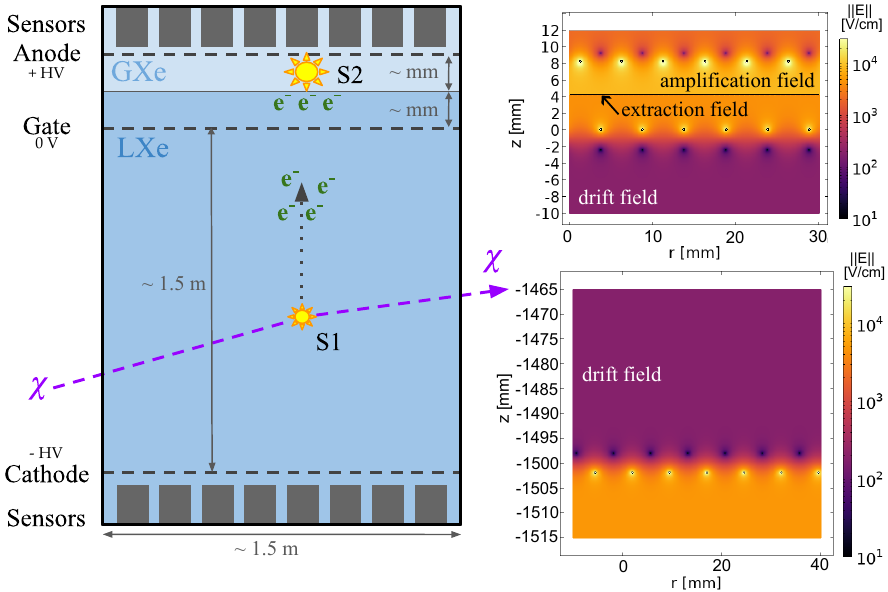}
\caption{Left: Diagram of a typical TPC \cite{nt_instrument,lz_detector}, showing the liquid and gaseous phases, the photo-sensors, the electrodes. 
In addition, the signal generation process is sketched out: An incident particle ($\chi$) scatters off Xe atoms, leading to an immediate scintillation signal (S1) and free electrons (e$^{-}$) through ionization. The free electrons drift to the gaseous phase, where they are accelerated, generating a secondary scintillation pulse (S2). Right: Typical field strength defined by the electrodes, with the amplification, extraction, and drift field regions labeled.}
\label{fig:tpc_sketch}
\end{figure}

A typical xenon dual-phase TPC consists of a large cylindrical volume filled with liquid xenon (LXe) with a smaller gaseous layer (GXe) in the top part \cite{nt_instrument, lz_detector, XLZD:2024gxx}.
The top and bottom ends of the TPC are equipped with arrays of photomultiplier tubes (PMTs).
Throughout the TPC, the electric fields are spanned by three electrodes:
the negatively charged cathode at the bottom of the TPC and the gate electrode positioned right below the interface between liquid and gas span the so-called drift field $\vec{E}_{\text{drift}}$.
The positively biased anode is situated in the gaseous phase and together with the gate defines the extraction and acceleration fields. 
In addition, in some configurations so-called screening electrodes are introduced in front of the PMT arrays to protect the sensors from electric fields \cite{nt_instrument}.
A diagram of a TPC can be seen in \autoref{fig:tpc_sketch}. 

Energy deposition in the TPC leads to scintillation photons and ionization electrons.
The prompt scintillation light is collected by the PMTs as the so-called S1 signal.
The electrons liberated by ionization can recombine with a xenon ion, and the resulting light adds to the S1 signal.
The electrons which do not recombine drift along $\vec{E}_{\text{drift}}$ towards the liquid-gas interface.
When the drifting electrons reach the top of the detector, they are extracted into the GXe by the extraction field and then accelerated, leading to a secondary scintillation signal, called S2.
The overall size of both signals provides information about the recoil energy, and the relative strength between S1 and S2 discriminates between nuclear recoils and electronic recoils.
The depth of the event is reconstructed by the time difference between S1 and S2, and the horizontal position of the event is deduced by the pattern on the top PMT array produced by the S2.

Correctly modeling the S2 signal requires a mostly homogeneous extraction field.
Although small inhomogeneities can be corrected at the analysis level, abrupt changes caused, for example, by substructures on the electrodes are difficult to mitigate.
Furthermore, a higher overall extraction field equals a higher chance of extracting the ionization electrons into the gas phase, increasing the strength of the S2 signal, and thus improving energy resolution and discrimination power between electronic and nuclear recoils. 
To ensure almost 100\% extraction, a field in the gas near the surface of $E\sim\SI{10}{kV/cm}$ is required~\cite{prop-scint}.
Furthermore, the strength of the drift field must be optimized with respect to the recombination of electrons and Xe ions and the extent of drift within the TPC, seeking values of  $E_\text{drift}\sim\SI{290}{V/cm}$~\cite{XLZD:2024gxx}.

The main challenges of electrode design and fabrication are the stability against electron emission in high electric fields, formation of homogeneous fields (minimal sagging along the entire electrode) and simultaneously guaranteeing maximal transparency for the scintillation light.

\subsection{Design and Limitations of Electrodes for Time Projection Chambers}
\label{subsec:RnD_at_kit}

Commonly employed TPC designs use variations of  parallel-wire electrodes \cite{nt_instrument} or hexagonal mesh electrodes \cite{next100_electrode, xenon100_instrument}, or a combination of both \cite{1t_instrument}.
Additionally, there exist more novel designs such as woven-wire electrodes \cite{lz_electrodes}, or coated electrodes on a transparent substrate \cite{darkside20k}.
In this work, we focus on improving parallel-wire and hexagonal mesh electrodes. 
Their performance was studied in previous generations of experiments and showed significant limitations. 
Both designs have technical challenges to be solved for future detectors, in particular to guarantee minimal deformation, ease of installation and repair, and optimal manufacturing including quality control. 

First, spanned parallel wires deform the ring preferentially in one direction, thus increasing the sagging and affecting the uniformity of the electric field. 
In contrast, a hexagonal mesh provides a more uniform tensioning and mechanical load in all directions on the plane, thus guaranteeing a more homogeneous bulk field between the electrodes. 
Additionally, regarding the ease of installing the electrode, the common strategy for a parallel-wire electrode with hundreds of wires is cumbersome and error-prone \cite{lz_electrodes}, unlike the case for an etched mesh, as demonstrated in \cite{next100_electrode}. 
On the other hand, the etched mesh is monolithic and can hardly be repaired in case of imperfection after manufacturing or during handling.

Concerning the current manufacturing capabilities, single-piece etched meshes larger than 1.5\,m with micrometer-scale precision are challenging to produce and procure \cite{next100_electrode}. 
In contrast, high-quality wires with diameters on the order of \SI{100}{\micro\meter} for parallel-wire electrodes are readily obtainable. 
At the same time, the fabrication of a monolithic support frame for wire installation has also encountered challenges, even for a size of approximately 1.5\,m in diameter. 
Therefore, at this stage it remains open which electrode design is more advantageous in terms of scalability.

A further challenge for either type of electrode is the potential of spurious electron emission.
Features, such as sharp edges, introduced during the production or handling of an electrode, can lead to localized extremely high fields.
At such points, these field strengths can overcome the work function of the electrode material and release electrons into the detector \cite{FN_field_emission,adam_bailey_thesis_em,PandaX:2014mem}, in the worst case, causing electrical breakdown.
As an example of structural deficiency, we report the case of the XENONnT detector.
During its commissioning, a broken wire in one of its electrodes resulted in the inability to operate the TPC with a drift field greater than $\sim$\,\SI{23}{V/cm} \cite{nt_instrument}.
The detector remained functional and was successfully used for performing a variety of scientific searches \cite{xenonnt_wimp_sr0, xenonnt_sr0_lower, xenonnt_b8}.
However, without an upgrade of its electrodes to alleviate the underlying issue, the detector was not capable of reaching its design drift field and sensitivity \cite{xenonnt_sensi}.

The development of the next generation of LXe dual-phase TPCs like for the XLZD detector is associated with multiple challenges and, following the aforementioned problems, some of the crucial ones pertain to the design and production of suitable electrodes.
For this reason, electrodes, regardless of their shape, need to be carefully designed to ensure their structural integrity.
Additionally, they need to be inspected for the presence of abnormal features and their effect on the electrode performance. 

\section{Parallel-Wire Electrodes} \label{sec:parallel_wire}

Here we consider a scenario where a parallel-wire electrode is acting as an anode in a LXe dual-phase TPC.
The developed electrode follows an initial design of the XENONnT experiment \cite{nt_instrument, francesco_toschi_thesis} and consists of an array of stainless steel (SS) parallel wires. 
Each end of a wire was bent 90$^{\circ}$ individually and fixed to the SS electrode frame by a pin made from oxygen-free high conductivity (OFHC) copper ($>99.99\%$ pure).  
The frame is a 24-gon with the inner diameter of \SI{1.33}{m}. 
The anode electrode is subject to several stringent requirements.
To achieve field homogeneity, the separation between the wire centers is the same across the entire electrode, which is \SI{5}{mm}, resulting in a total of 265 wires for the entire electrode.
The gap between the anode and the liquid level being in the range from a few millimeters to a centimeter, significantly affects the amplification and width of the S2, leaving a small space for wire sagging.  
At the same time, the electrostatic force on the anode is about twice that on the gate due to the different dielectric constants of liquid and gaseous xenon. In addition, it is in the same direction as the gravitational force. Thus, the load on the anode wires is larger than that on the gate electrode. 
The requirements for a parallel-wire electrode as an anode are listed in \autoref{tab:anode_requirement}.
\begin{table}[htb!]
\caption{Requirements for a parallel-wire electrode as an anode. The parameter $A$ is the cross-sectional area of the wire. 
The thermal stress ($\sigma_s$) was calculated at \SI{75}{MPa} assuming the Young's modules of the wire to be \SI{200}{GPa} at \SI{175}{K} \cite{coefficient_thermal_expansion}, the coefficient of thermal expansion at \SI{1.5e-5}{} \cite{coefficient_thermal_expansion}, and the temperature difference across the electrode at \SI{25}{K}.
The yield tensile strength ($\text{YTS}_{0.01\%,\,\text{meas.}}$) was measured as described in \autoref{sec:wire_test}. 
The definition of the showcased parameters $h_{\text{max}} $ and $F_{ax}$ is provided in the text below.
} \label{tab:anode_requirement}
\begin{center}
\begin{tabular}{l l l}
Quantity & Symbol & Requirement\\
\hline
Frame deformation & & $<$\,\SI{5}{mm}\\
Sagging limit & $h_{\text{max}} $& $<$\,\SI{0.5}{mm} \\
Safety factor & $\text{YTS}_{0.01\%,\,\text{meas.}}\,/\,(\sigma_s+F_{ax}/A)$& $ \gtrsim 1.5$ \\
Transparency & & $\gtrsim 90\%$\\
\end{tabular}
\end{center}
\end{table}

\begin{figure}[htb]
\includegraphics[width=\textwidth]{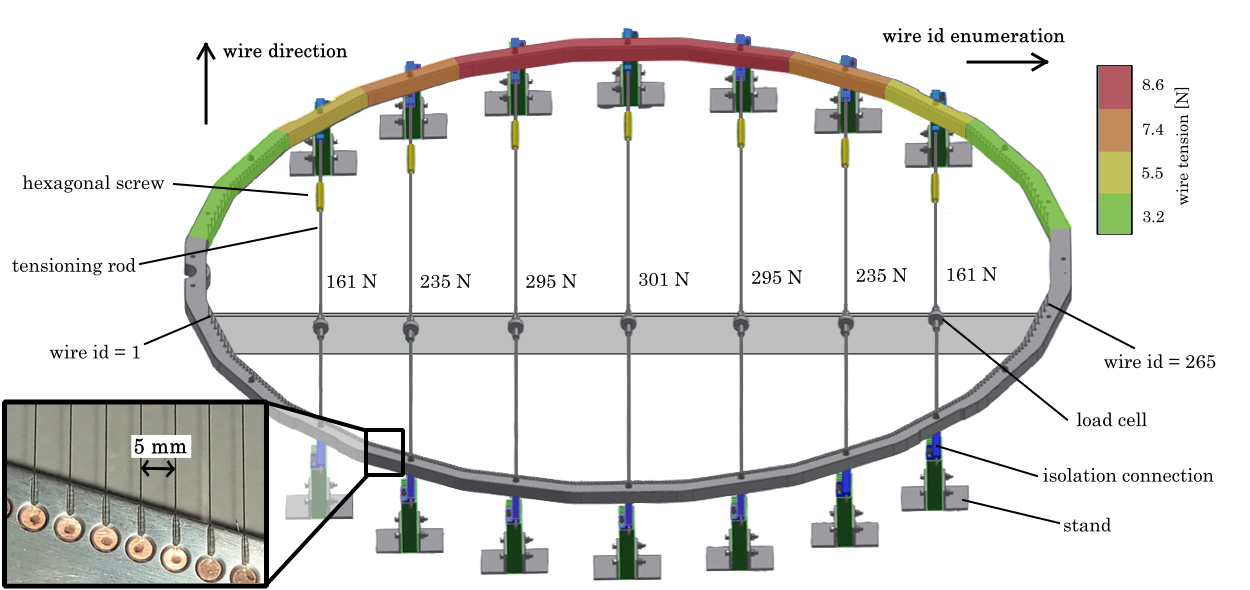}
\caption{Schematic drawing of the wire electrode frame mounted onto the tensioning system. The color on the electrode frame indicates the target tension of the wires in each section. The tensioning system underneath the frame consists of the tensioning rods that hold the ring, with a hexagonal screw (in yellow) adjusting the tension, and the load cells that measure the axial tension on the respective tensioning rod. 
The required tension on the tensioning rods is shown to the right of each rod.
The band underneath the load cells limits the lateral load on the load cells. The stands support the entire structure, and the isolation connections allow the movement and relaxation of the ring and the tensioning rods. 
The inset shows a picture of the assembled electrode, where individual wires are separated by \SI{5}{mm} and fixed in place by copper pins. 
}
\label{fig:tensioning}
\end{figure}

To reduce sagging of the wires during detector operation, axial tension was applied to each wire during assembly. For long wires with small sagging, the load along the wire can be assumed to be uniform. Thus, we estimate the axial tensions $F_{ax}$ on the wires from the sagging model in \cite{RoarkSS2022} :
\begin{equation}\label{eq:sagging_uniform}
    F_{ax} = \frac{(\omega_g + \omega_{el}) L_0^2}{8h_{\text{max}}}
\end{equation}
where $F_{ax}$ is the axial tension. $\omega_g$ and $\omega_{el}$ are the respective unit loads due to gravity and electrostatic force. $L_0$ is the undeformed length of the wire between supports. $h_{\text{max}}$ is the maximum sagging. 
The contribution from the elasticity of the wire was found to be less than $0.2\%$, and was therefore neglected \cite{wire_with_elasticity, francesco_toschi_thesis}.
We calculated $\omega_{el}$ using a 2D axially symmetric field simulation, considering a specific field configuration in the detector. 
Buoyancy is neglected in the load as it is four orders of magnitude smaller, given that the anode is in GXe.
For a wire with a diameter of \SI{0.216}{mm}, we estimated $(\omega_g + \omega_{el})$ to be \SI{1.81e-5}{N/mm} downwards.
The undeformed length of the longest wire is \SI{1340.4}{mm}, and the maximum sagging was targeted at \SI{0.5}{mm}.
Although the tension is proportional to $L_0^2$, we assort the wires into four groups of different tension values from 3.2\,N per wire for the shortest ones up to 8.6\,N per wire for the longest ones to ease the assembly process, as showcased in \autoref{fig:tensioning}. 
Due to the low tolerance on sagging and subsequently high axial tension required, we performed additional simulations and tests to investigate the structural integrity of the electrode.

\subsection{Mechanical Simulations of the Electrode Frame} \label{sec:mech_sim_frame}

The frame is constructed from a stainless steel SS316 slab with a yield tensile strength (YTS) of approximately \SI{240}{MPa} \cite{SS_book}. 
Previous experience showed that the assembled electrode tended to release internal stress after a cooling cycle, leading to increased sagging of the wires \cite{francesco_toschi_thesis}. 
To avoid such relaxation of the frame, it was annealed before the final machining to relax its internal mechanical stress.

We performed mechanical simulations to determine the stress and deformation of the electrode frame due to the aforementioned axial tensions of the 265 wires.
For this, we used ANSYS Workbench 2021 for finite element analysis (FEA).
Instead of having $265\times3$ force components as input, the forces on each section of the polygon were combined into a single force vector to simplify the calculation.
To further increase the computational efficiency, we applied mirror symmetry along the horizontal axis in \autoref{fig:tensioning}. 
The simulation setup resulted in 12 input forces for the FEA. 

To evaluate the component stresses on the electrode frame, the von Mises stress criterion was applied which is commonly used for ductile materials such as steel. 
The multiaxial stress state in the real component was represented by an equivalent uniaxial stress, providing a comparable measure of the material stress. 
This equivalent stress value can then be compared to the material's YTS to assess the risk of material failure.

With iterations of the simulation, the geometry of the electrode frame was optimized to optimally withstand the stress of the wires. 
The height of the SS frame was increased to \SI{24}{mm} compared to the original XENONnT design of \SI{18}{mm}, and the width varies along the circumference from 30.975~mm to 33.975~mm. 
The final maximum stress was 98.6~MPa, where the cross-section is the smallest. Compared to the YTS of the material, this design secured a safety factor of 2.4 against permanent plastic deformation.
In addition, given the load of the wire tension, the ring should deform by a total of 4.41 mm outwards perpendicular to the wire direction.
Parallel to the direction of the wires, the ring deformation is expected inwards by 4.44 mm in diameter.
Based on this optimized design, the raw pre-manufactured ring was annealed, the final frame was then manufactured and electropolished by Mühlbauer Parts \& System (MPS) \cite{MPS}.

\subsection{Wire Tensile Tests at Different Temperatures} \label{sec:wire_test}

There are several contributions to the stress on a wire at different stages.
First, when installing the wire to the frame with copper pins, each wire experienced a highly geometry- and method-dependent stress, which cannot be properly estimated. 
Second, the wire was subjected to axial tension to reduce sagging, as indicated in \autoref{fig:tensioning}.
Then, when filling the TPC with LXe at $\sim$\,\SI{177}{K}, 
the electrode frame cools down slower than the wire due to their surface-to-volume ratios, inducing thermal stress on the wire.
The thermal stress amounts to \SI{75}{MPa} at a temperature difference of \SI{25}{K}, resulting in an additional force of $\sim$\,\SI{2.75}{N} for a wire with diameter of \SI{0.216}{mm}.
During the operation, there is no thermal stress anymore. 
The electrostatic force between the electrodes remains. 
However, its contribution, together with the gravitational load, is an order of 100 times less than the axial tension, thus negligible. 
Therefore, the maximum stress considered was the sum of the largest axial tension and the thermal stress, estimated to be \SI{11.35}{N} for a wire with a diameter of \SI{0.216}{mm}.

\begin{figure}[htb!]
\centering
\includegraphics[width=1.0\textwidth]{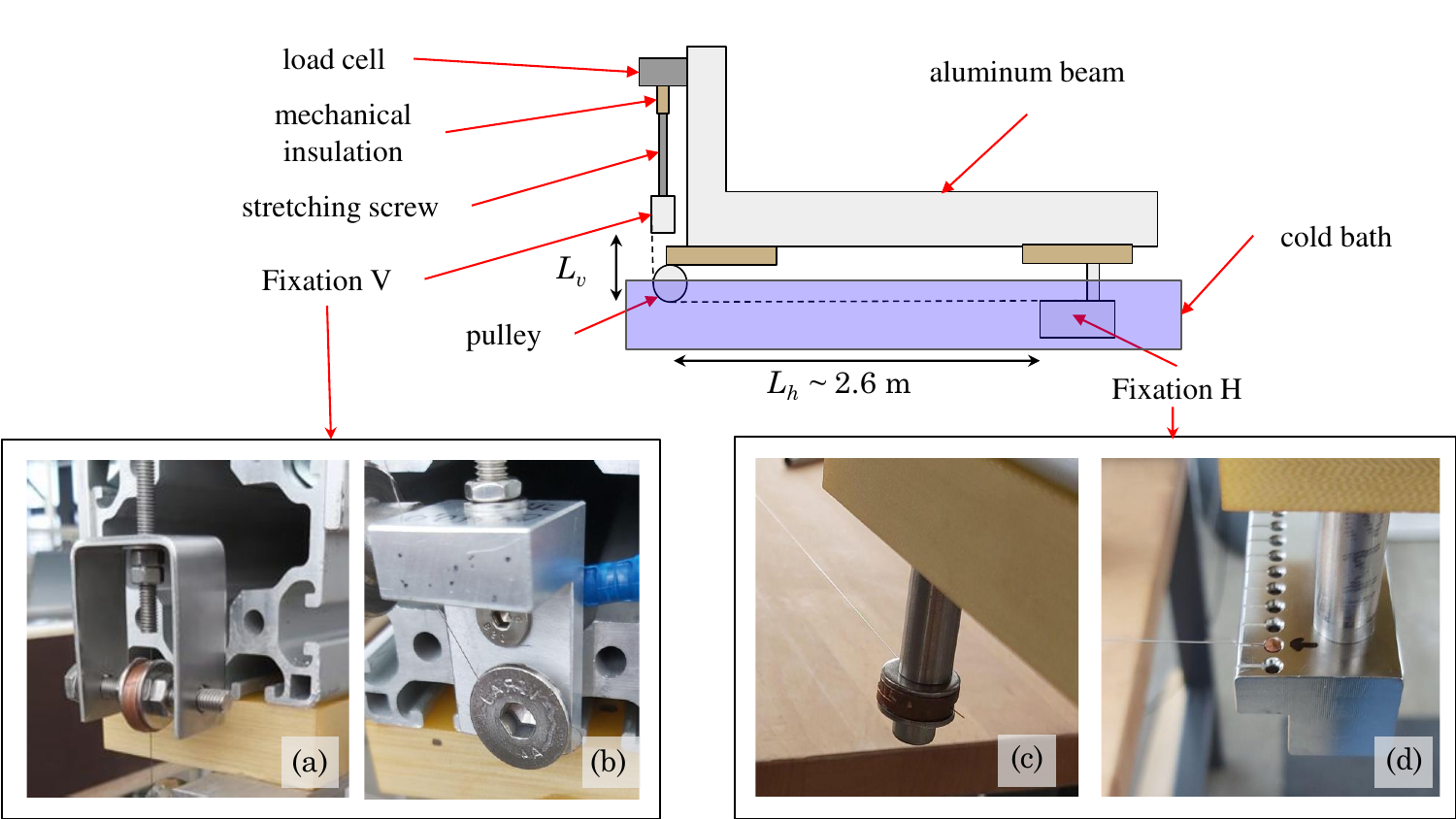}
\caption{Schematic drawing (not to scale) and photos of different components of the wire test setup. The dashed line indicates the wire sample. 
The L-shaped aluminum beam profile is supported externally outside the cold bath, which is not indicated in the drawing. 
The vertical part of the setup consists of Fixation~V that fixes the wire sample, a stretching screw that was rotated manually to increase the strain of the wire sample, a load cell measuring the axial tension on the wire sample, and the mechanical insulation that decouples the movement of the stretching screw and the load cell. 
Two types of Fixations~V were used, either a copper gasket (a), or a screw fixing (b). 
The wire passed through the pulley to enter the horizontal part of the setup. 
Fixation~H fixed the wire sample to the end of the horizontal part, either by two copper gaskets clamping the wire (c) or by a mock-up of the electrode frame segment where a copper pin was inserted to fix the wire in place (the copper pin fixation) (d). 
$L_v$ and $L_h$ represent the vertical and horizontal length of the wire, measured from the centre of the pulley.}
\label{fig:wire_test_setup}
\end{figure}

To search for a suitable wire candidate withstanding these conditions, a custom-made tensile test setup, as shown in \autoref{fig:wire_test_setup}, was built to perform measurements in a temperature-controlled environment. 
We performed tensile tests with several wire candidates, summarized in \autoref{tab:wire_can}, at 6 temperatures from \SI{200}{K} to room temperature. 
We also performed the same test with an electrode frame mock-up and the copper pin, later referred to as the copper pin fixation, to see the additional effect of the fixation itself.

\begin{table}[h!]
\caption{Data sheet information for the candidates of the wire tests. YTS values refer to an offset of 0.2\%. Measurements for ultimate tensile strength (UTS) and YTS are presumably taken at room temperature.} \label{tab:wire_can}
\begin{center}
\begin{tabular}{l c c c c }
Manufacturer        & Diameter {[}mm{]}  & Material (AISI)              & UTS {[}MPa{]}          & \begin{tabular}[c]{@{}l@{}}YTS {[}MPa{]}\end{tabular} \\ \hline
\hline
\begin{tabular}[c]{@{}l@{}}California Fine \\ Wire (CFW) \cite{CFW}\end{tabular} & {0.216} & {316 annealed}   & {884}       & {725}             \\ \hline
Vogelsang \cite{vogelsang} & {0.212} & {316 semi-hard}  & {1000-1200} & {N.A.}            \\ \hline
Dahmen \cite{Dahmen}               & {0.221} & {316L full-hard} & {1916.5}    & {1528.5}          \\ 
\end{tabular}
\end{center}
\end{table}

The setup consists of an L-shaped aluminum beam profile, on which the wire sample was fixed by \emph{Fixation~H} on the horizontal end and \emph{Fixation~V} on the vertical end.
The length of the horizontal and vertical parts of the wire is approximately \SI{2.6}{m} and of the order of \SI{10}{cm} respectively, as measured for each measurement.
For Fixation~V, we used either copper gaskets or a screw clamp to account for different systematic errors.
Nevertheless, there were no significant differences observed between the two Fixations~V used.
For Fixation~H, we used either copper gaskets or an electrode frame mock-up with a copper pin.
Fixation~H was attached to a screw system, which was used to stretch the wire sample in the axial direction.
A \SI{200}{N} load cell positioned at the top of the vertical part measured the axial force on the system.
The load cell was calibrated periodically, and the average values were used for the analysis. 
Note that the wire samples and the load cell did not rotate together with the screw system.

For each measurement, we first mounted a wire sample into the setup.
The cold bath was prepared separately by adding dry ice to a bath of ethanol inside the partially thermally isolated bath. 
The liquid level was set to cover approximately half of the pulley.
Two PT100 temperature sensors were used to monitor the temperature at different positions of the cold bath.
As the temperature approached the desired value, we repositioned the entire L-shaped setup so that the horizontal part of the wire was completely submerged in the cold bath. 
A data acquisition module from National Instruments \cite{national_instruments} continuously read out the data from the load cell and the temperature sensors during a measurement at the rate of \SI{50}{Hz}.
The stretching screw was turned in steps of one-fourth of a turn while the stress was recorded by the load cell.

For each measurement, we obtained a stress-strain curve.
The linear part of the stress-strain curve is defined as the elastic range of a material.
A linear function can be fitted to the elastic range, where the slope is defined as the Young's Modulus. 
The YTS is defined as the stress at which the material starts to deform permanently. 
The value can be estimated by the intersection point between the stress-strain curve and the linear fit shifted by a certain amount. 
The conventional shift, or offset, is 0.2\%, measuring the stress where the material is permanently deformed by 0.2\%. 
To fullfil the more stringent requirement of sagging, we used an offset of 0.01\% instead.

To test the reliability of the pin fixation after a thermal cycle, we performed an additional tensile test after the system underwent a thermal cycle in a \SI{200}{K} cold bath, which is close to the detector operating temperature of \SI{175}{K} \cite{nt_instrument}.
We cooled down the setup for 1 minute until the measured temperature around the wire, monitored by a PT100 sensor, stabilized at \SI{200}{K}.
Then, we took out the setup to warm it up until the temperature around the wire went back to room temperature before the tensile test.
In case of a significant thermal relaxation between the copper pin fixation, the wire could slip, resulting in a reduction in the measured YTS and ultimate tensile strength (UTS) values, which was not observed.

\begin{figure}[htb!] 
\begin{center}
\includegraphics[width=0.7\textwidth]{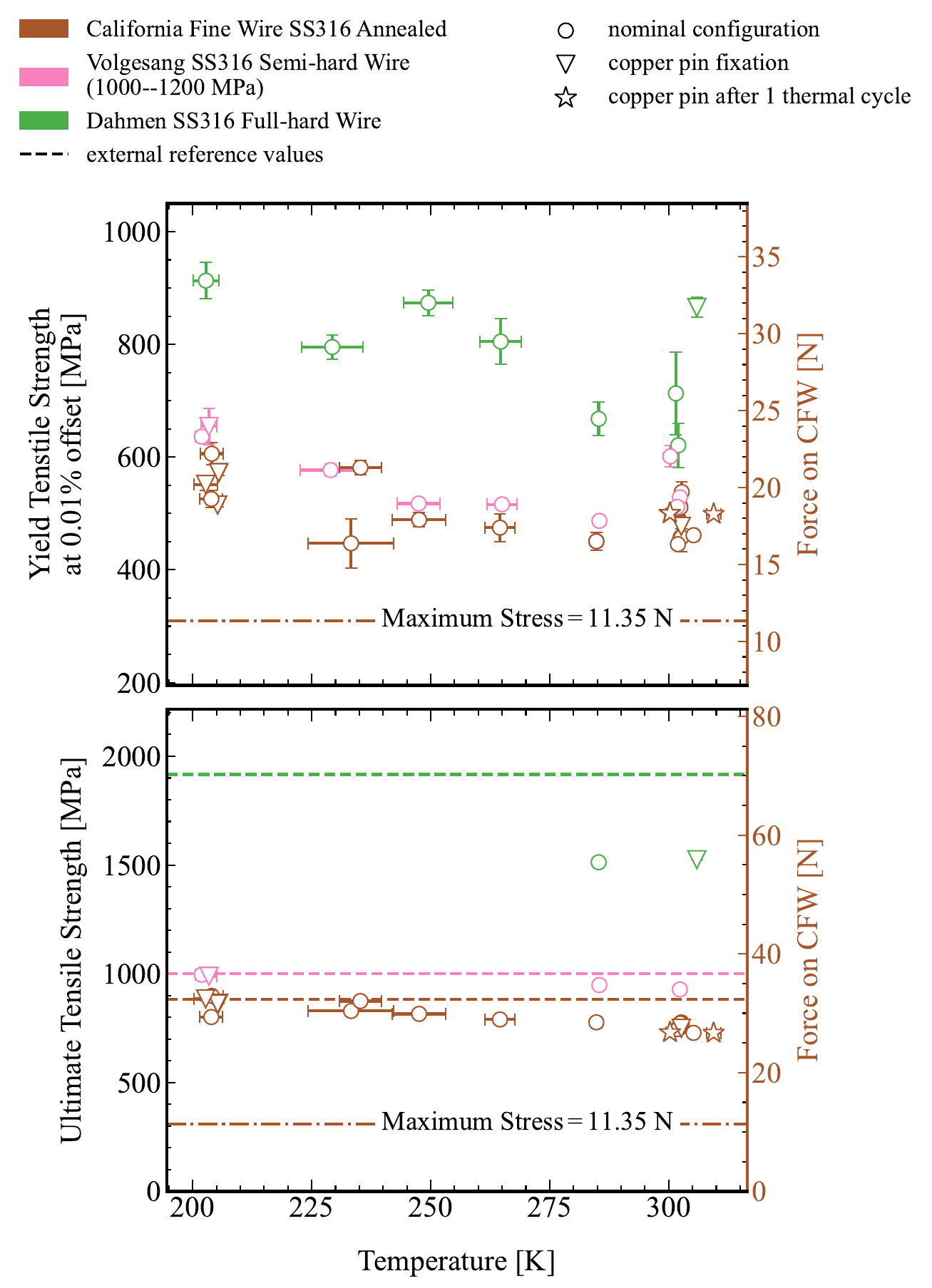}
\end{center}
\caption{
Results from the tensile tests at various temperatures. 
The top plot shows the yield strength at 0.01\% offset, meaning the wire undergoes 0.01\% additional deformation beyond the elastic range. 
The bottom plot shows the ultimate tensile strength of the sample. 
The right axes on both plots indicate the conversion to force acting on a wire with 0.216\,mm diameter as the CFW sample.
The colors indicate the wire samples. 
The different markers represent the different test configurations.
Circles: tests with the copper gasket fixation (\autoref{fig:wire_test_setup}(c)). 
Inverted-triangles: tests with the copper pin fixation, where the wire was bent by 90$^{\circ}$ (\autoref{fig:wire_test_setup}(d)). 
Stars: tests with the copper pin fixation and underwent 1 thermal cycle. 
Dash-dotted line: the expected maximum stress on the longest wire, including the stress due to tensioning and thermal shrinkage when cooling down the TPC. The temperature gradient was assumed to be \SI{25}{K} based on internal communication. 
Dashed lines: the UTS values from the data sheet, as shown in \autoref{tab:wire_can}.
}
\label{fig:wire_test_result}
\end{figure}

The resulting values of YTS for an offset of 0.01\% as well as the UTS at different temperatures are shown in \autoref{fig:wire_test_result}.
The temperature was the average value recorded by the two PT100 sensors inside the bath.
The error bound represents the maximum and minimum temperatures recorded by the sensors throughout each measurement.
For each measurement, the YTS and its error were obtained from linear regressions performed over different fitting ranges, with the ranges chosen to ensure $R^2 > 0.99$. 
The length of the vertical part of the wire, which was not submerged in the cold bath, was checked to be an insignificant error compared to the fitting error.
The error on the change in length and rolling friction was also estimated to be negligible.
We validated the measurement by comparing the YTS at 0.2\% offset to the data sheet values. 
For CFW, the mean value we measured at room temperature is consistent with the datasheet value, being $744\pm12$\SI{}{MPa} and \SI{725}{MPa} respectively. 
The error bar is the standard deviation of the measurements to account for systematics between configurations. 
For the Dahmen wire, the 0.2\% offset at room temperature was out of the operating range of our setup.
A datasheet value was not available for the Vogelsang wire sample.
The UTS was determined as the maximum stress in the stress–strain curve for each measurement, provided that the wire sample broke.
Similarly to the YTS measurement, for some wire samples from Dahmen and Volgesang, breakage did not occur within the maximum strain achievable with the experimental setup.

From our results, all wire samples tested can withstand the maximum stress, caused by the axial tension and the thermal stress, with less than 0.01\% plastic deformation.
Specifically, the maximum stress for the longest CFW wire is \SI{11.35}{N}, indicated by the dash-dotted line in \autoref{fig:wire_test_setup}.
Mechanical simulations indicate that under the gravitational and electrostatic forces, its sagging would subsequently be less than \SI{0.5}{mm}.
Compared to the estimated maximum stress, the average safety factor for CFW until deformation is $1.6\pm0.2$.
Additionally, as the temperature decreased to \SI{200}{K}, we observed a slight increasing trend in the YTS at a 0.01\% offset, as suggested in \cite{lz_electrodes}, with correlation coefficients ranging from $-0.62$ to $-0.68$ for all wire candidates.
The YTS at a 0.01\% offset with copper pin fixation differs by less than 5\% compared to the copper gasket fixation, regardless of whether the system underwent a thermal cycle.

Apart from YTS and UTS, there are additional metrics to consider when selecting suitable candidates, namely the ductility and the surface treatment. 
Specifically, the Dahmen wire is prone to breaking during installation due to the stress from the copper pin fixation.
Although the YTS is higher, hard wires are more brittle than other wires. 
Annealed SS, on the other hand, exhibits enhanced ductility among the three candidates, providing a longer range of plastic deformation before fracture.
This property is ideal to avoid wire breakage, which could result in electrode short-circuiting or sparking at high voltages. 
Since CFW fulfills the mentioned requirements and has been tested in the XENONnT experiment, we subsequently used the CFW sample for the remaining tests. 
With a pitch of 5\,mm and the wire cross section of 0.216\,mm, an optical transparency of 95.7\% is obtained, fulfilling the requirement stated in~\autoref{tab:anode_requirement}.

\subsection{Wire Installation Procedure}

In previous experiments, the common strategy of mounting several hundreds of parallel wires onto an electrode frame was a challenging and iterative process \cite{lz_electrodes, mannino_thesis}.
For each newly installed wire, the electrode frame deforms simultaneously, altering the tension of those wires already mounted.
For an even larger electrode, a new installation method is required to reduce the number of individual steps, the time needed, and potential inaccuracies during the process.
In this work, we developed a new set of tools and procedures for a robust parallel-wire electrode installation, as shown in \autoref{fig:tensioning}. 
Essentially, we deformed the electrode frame to the final equilibrium shape before mounting the wires. 
Thus, the frame held its shape throughout the wire installation. Afterwards, the assembled electrode was dismounted from the station.

The detailed procedure is as follows. 
First, seven \SI{5}{mm}-diameter tensioning rods and stands are prepared on a table with a rough positioning relative to the electrode frame, parallel to the wire direction.
On each rod, there was a brass hexagonal key for adjusting its length.
Each rod also had a load cell for measuring the force.
Second, the electrode frame was mechanically coupled to the tensioning rods with screws.
Since the stands had the freedom to move along the rod's direction and horizontally on the table, the frame and the tensioning rods were decoupled from external anchors except through the frictional force.
Then, by reducing the length of the tensioning rods with the hexagonal screw, the electrode frame was pulled into the final equilibrium shape as if all the wires were mounted. 
The required tension on each rod is calculated by FEA, as shown in \autoref{fig:tensioning}.
When tuning one rod, the tension on other rods also changed non-linearly, as in the case when installing wires.
However, this procedure reduced the complexity from optimizing the tension on the 265 wires and the ring to only seven tensioning rods, which took around half a day to reach the required values within $\pm$\SI{10}{N} altogether.
After tensioning the frame, the whole system was kept unchanged for 2 days to relax before installing any wires. 

During wire installation, the shape of the electrode frame underwent limited change due to the stiffness of the tensioning rods, amounting to $<0.35\pm0.24$\,\SI{}{mm}.
The wires were installed with their target tension values as indicated in \autoref{fig:tensioning}. 
With custom-made tools, we tensioned the wire on both ends through the pin hole and fixed the wire using copper pins. 
All individual wire tensions were measured by a laser system pioneered in \cite{Prall:2009ep_laser_system}, updated and fabricated by \cite{Jens2023}. 

After the installation of all wires, assuming the tensioning rods to be perfect rigid bodies, all the load on the tensioning rod should be transferred to the tensioned wires, holding the frame in its final shape in equilibrium.
In reality, a small net force component remained between the tensioning rods and the frame which was eventually released through screws at the end of each rod.
Finally, the frame was dismounted from the tensioning system. 

\subsection{Test Installation}

\begin{figure}[htb]
\includegraphics[width=\textwidth]{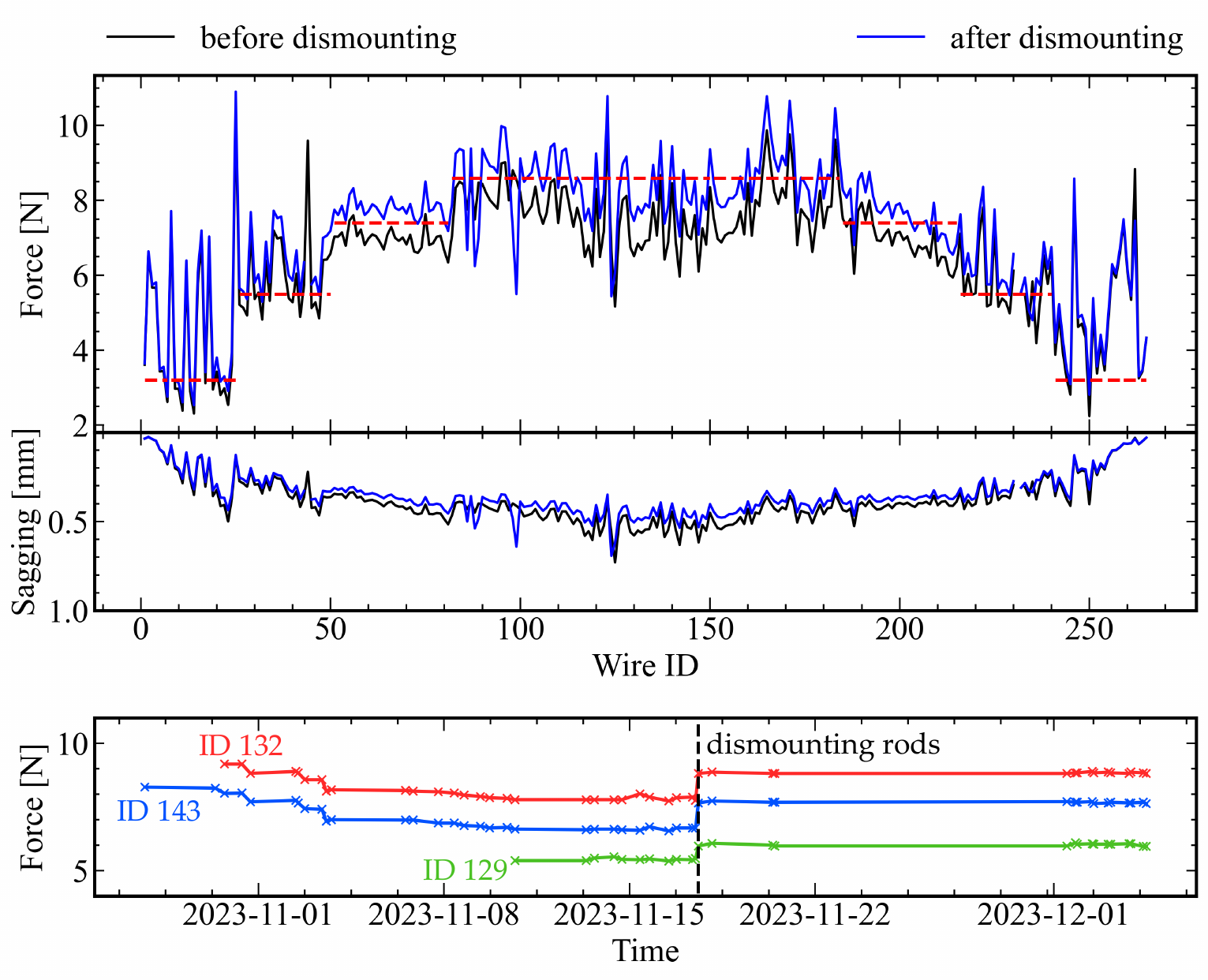}
\caption{Top panel: Wire tension and resulting expected wire sagging across all wire IDs before (black) and after dismounting (blue) of the tensioning rods. 
The wire ID notation is defined in \autoref{fig:tensioning}.
The red dashed lines indicate the target tension of the wires.
Bottom panel: Wire tension over time of selected wires. 
The vertical dashed line indicates the time when the tensioning rods were dismounted.}
\label{fig:wire_tension}
\end{figure}

We performed a test installation in a non-cleanroom environment with the aforementioned procedures.
At the same time, we monitored the tension of all mounted wires and all the tensioning rods twice a day. 
We also measured the wire tensions of the fully assembled electrode right before and after dismounting the electrode from the tensioning system. 
\autoref{fig:wire_tension} shows the resulting wire tensions from this test installation, their expected sagging during the operation in a TPC with field configuration as in \autoref{fig:tpc_sketch}, and the force evolution of a few selected wires during and after the tensioning procedure. 
As can be seen, the final tension is centered around the target value (red dashed lines) both before and after dismounting the electrode from the tensioning system. 
The computed sagging of the longest wires located in the central section averaged at \SI{0.51}{mm} as expected, with a maximum of \SI{0.67}{mm}.

The deformation of the frame was measured by a \SI{4}{m} long 7-axis FARO\textsuperscript{\textregistered} Quantum$^\text{M}$ FaroArm\textsuperscript{\textregistered} \cite{faro_arm, faro_arm_tech_sheet} as listed in \autoref{tab:faro_arm_measurement}. 
The values are consistent with the FEA simulation within the measurement uncertainty of $\pm$\SI{0.24}{mm}. 
This uncertainty is given by the different surface points used to obtain the measurements. 
After all the wires are installed, the frame deformed slightly upward along the wire direction and slightly downward across the wire direction, thus the unevenness of the frame increased from \SI{0.43}{mm}, with no wires installed, to \SI{0.73}{mm}, with all wires installed and the rods dismounted. 

\begin{table}[h]
    \caption{Deformation of the electrode frame derived by FEA simulation and from measurement with FaroArm\textsuperscript{\textregistered} at different stages of the assembly. The deformation parallel and perpendicular to the wires is shown in the second and third columns, respectively. All measurements with the Faro Arm have an uncertainty of \SI{0.24}{mm}. }
    \centering
    
    \begin{tabular}{l | c c}
     & \multicolumn{2}{c}{\textbf{Deformation [mm]}} \\
    Status of the Frame & $\parallel$ to wires & $\perp$ to wires \\ 
    \hline
    \hline
    \textbf{Only Tensioning Rods installed}    &              &                             \\ 
    simulation & 4.46 & -4.48 \\
    measurement & 4.26 & -4.34 \\
    diff. (simulation - measurement) & 0.20 & -0.14  \\
    \hline
    \hline
   \textbf{Fully assembled and rods dismounted}    &              &                              \\ 
   simulation & 4.44 & -4.41 \\
   measurement & 4.48 & -4.35 \\
   diff. (simulation - measurement)& -0.04 & -0.06 \\
    \end{tabular}

    \label{tab:faro_arm_measurement}
\end{table}

As can be seen from \autoref{fig:wire_tension}, there are significant fluctuations of the individual wire tensions around the target value. 
In this test setup, we realized as origin of these fluctuations an oxide layer on the copper pins altering the wire tension during the pin insertion process. 
Later installation with cleaned pins significantly reduced the difference between the realized and the target tension.
In addition, since the tensioning rods are not perfectly rigid, the tension on the wires was reduced as more wires were inserted. 
Examples are shown in \autoref{fig:wire_tension}~(bottom). 
The earlier the wire was installed, the larger the reduction, with a maximum drop of $\sim$\,\SI{2}{N}. 
Since the installation started from the middle of the electrode towards the sides, the tension of most wires in the middle sections fell slightly below the target value which was recovered after releasing the entire electrode. 
This intermediate effect could be reduced by using thicker tensioning rods to increase the stiffness of the tensioning system as well as mounting the wires from the outermost positions towards the center.

The tension for all wires increased after dismounting the electrode from the tensioning system, with a maximum of \SI{1.1}{N}.
This can be explained by the non-zero total pulling tension of around \SI{300}{N} (18\% of the initial tensioning force) remaining on the tensioning rods after all wires were installed.
As a result, the electrode frame extended slightly backward and stretched the wires after being released from the tensioning rods.

The conducted tests demonstrate a reliable electrode design with a new installation method reducing the complexity of installing a parallel-wire electrode. 
However, a simple extrapolation to a next-generation electrode with 2.6--3\,\SI{}{m} diameter would not be feasible. 
As an example, for such large wire lengths, the requirement of a maximal sagging of 0.5\,mm would lead to tensions beyond the YTS of the tested wires such as CFW. 
Either other materials or larger wire diameters could overcome this limitation, eventually compromising other requirements such as the optical transparency of the electrode. 
Another limiting fact is the fixation of the wires by pins perpendicular to the wire direction. 
This not only limits the usage of harder but more brittle SS wires with higher YTS but also leads to a z-component of force on the frame, resulting in a saddle-shaped profile after full installation only avoidable with a thicker frame design.

\section{Hexagonal Mesh Electrodes}\label{sec:hex_mesh}

\begin{figure}[htb]
\centering
\includegraphics[width=\textwidth]{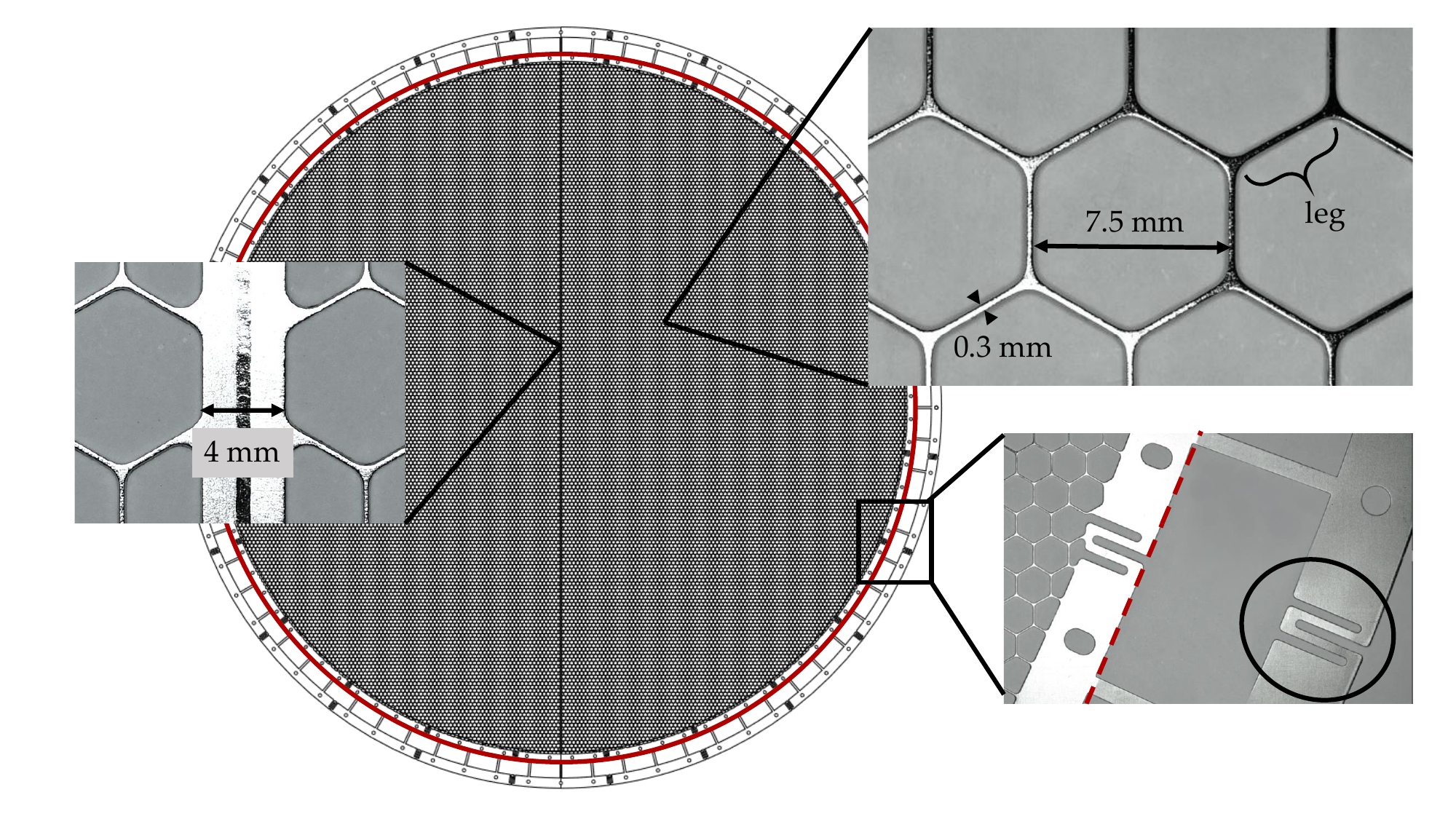}
\caption{Illustration of the hexagon mesh with insets showcasing close-up images of several locations on an electropolished mesh.
The leftmost inset shows the \SI{4}{mm} wide laser-welded strip and the welding seam in its middle. 
The top-right inset showcases the dimensions of an individual hexagon in the mesh with leg thickness of \SI{0.3}{mm} and opening of \SI{7.5}{mm}.
The bottom-right inset depicts the outer edge of the mesh, with the inner and outer rings of the mesh intended for stretching and fixation on the mesh on the electrode frame, respectively. 
An omega-shaped section on the outer ring is circled in black.
The red line indicates the boundary beyond which the structure is intended to be mechanically cut after installation of the mesh on the electrode frame.  
}
\label{fig:hex_mesh}
\end{figure}

Etched meshes with hexagonal-shaped openings and a typical thickness $\sim$\,\SI{100}{\micro\meter} are commonly used as electrodes in a variety of TPCs \cite{next100_electrode, 1t_instrument}.
However, due to challenges associated with manufacturing whole meshes at a scale of $\sim$\,\SI{1.5}{\meter}, as required for detectors such as XENONnT \cite{nt_instrument}, we chose instead a semi-modular approach.
As depicted in \autoref{fig:hex_mesh}, several SS304 half-meshes with an inner ring radius of \SI{1394}{mm} were produced via photochemical etching at PCM Products, Inc. \cite{PCM}, and then laser-welded together at KIT. 

The meshes were produced with openings measuring \SI{7.5}{mm} and leg thickness of \SI{0.3}{mm}.
The choice of leg thickness was driven by exploratory electropolishing which was conducted with test meshes with leg cross-section of \SI{0.2}{mm} at OTG Oberflächentechnik Gronau \cite{OTG}.
After electropolishing these test meshes exhibited loss of material at the outer radii which was significant enough to cause breaks in individual hex legs.
Therefore, to minimize the number of potentially cut hex legs, we opted for thicker mesh legs of \SI{0.3}{mm}.
Additionally, attempts were made to investigate the potential use of water-jet cutting as an alternative to photochemical etching.
However, preliminary results suggested that photochemical etching was superior in terms of production time, material use and the resulting mesh quality.

Each of the half-meshes has an outer and an inner ring for stretching and fixation of the mesh on the electrode frame, respectively.
The rings are interconnected via narrow strips, which are intended to be mechanically severed after the mesh is stretched and fixed to the electrode frame.
Additionally, omega-shaped relief sections are located at regularly spaced gaps along both the inner and outer rings, facilitating greater flexibility of both rings.
Four of the produced half-meshes were laser-welded along a dedicated \SI{2}{mm} joining edge on each half, resulting in a central welded strip of \SI{4}{mm} across the final two whole meshes.
The strength of the connection produced by the laser welding was assessed in a series of tests that will be further elaborated on in \autoref{sec:defects_features}.
A ring-shaped SS electrode frame with an outer diameter of \SI{1395}{mm} and thickness of \SI{20}{mm} was produced and electropolished by MPS~\cite{MPS}.
To host the electrode mesh, the ring has a semi-circular cross-section and features 96 threaded M4-sized holes through which the mesh could be fastened to the frame.

\subsection{Mechanical, Optical \& Electrostatic Simulations}\label{sec:hex_mesh_sims}

Simulations were performed to quantify the effect of the central welded strip on the electric field homogeneity. 
Additional simulations were conducted to ensure the mechanical integrity and to assess the reduction in optical transparency that results from choosing a hexagonal mesh over the parallel-wire design.

Given the high field between the anode and the gate in the gaseous phase, inhomogeneity in the electrode's geometry largely affects the S2 signal. 
Therefore, the design with the central welded strip is not the most suitable candidate for the anode and the gate electrodes.
The subsequent discussion then focuses solely on the cathode electrode with its requirements as stated in~\autoref{tab:cathode_requirement}.
To study the effect of the \SI{4}{mm} welded strip on the field homogeneity, we performed both a full 3D and a local 3D electrostatic field simulation.
We used the KEMField package within the open-source software Kassiopeia \cite{Kassiopeia} for the full 3D simulation.
The package uses the boundary element method for the surface charge calculation and the fast multipole method to perform the field calculation.
Since the hexagonal geometry is computationally expensive, we used wire grids together with a \SI{4}{mm} wide plate on top of the grid to represent the electrode. 

\begin{figure}[htb!]
\centering
\includegraphics[width=\textwidth]{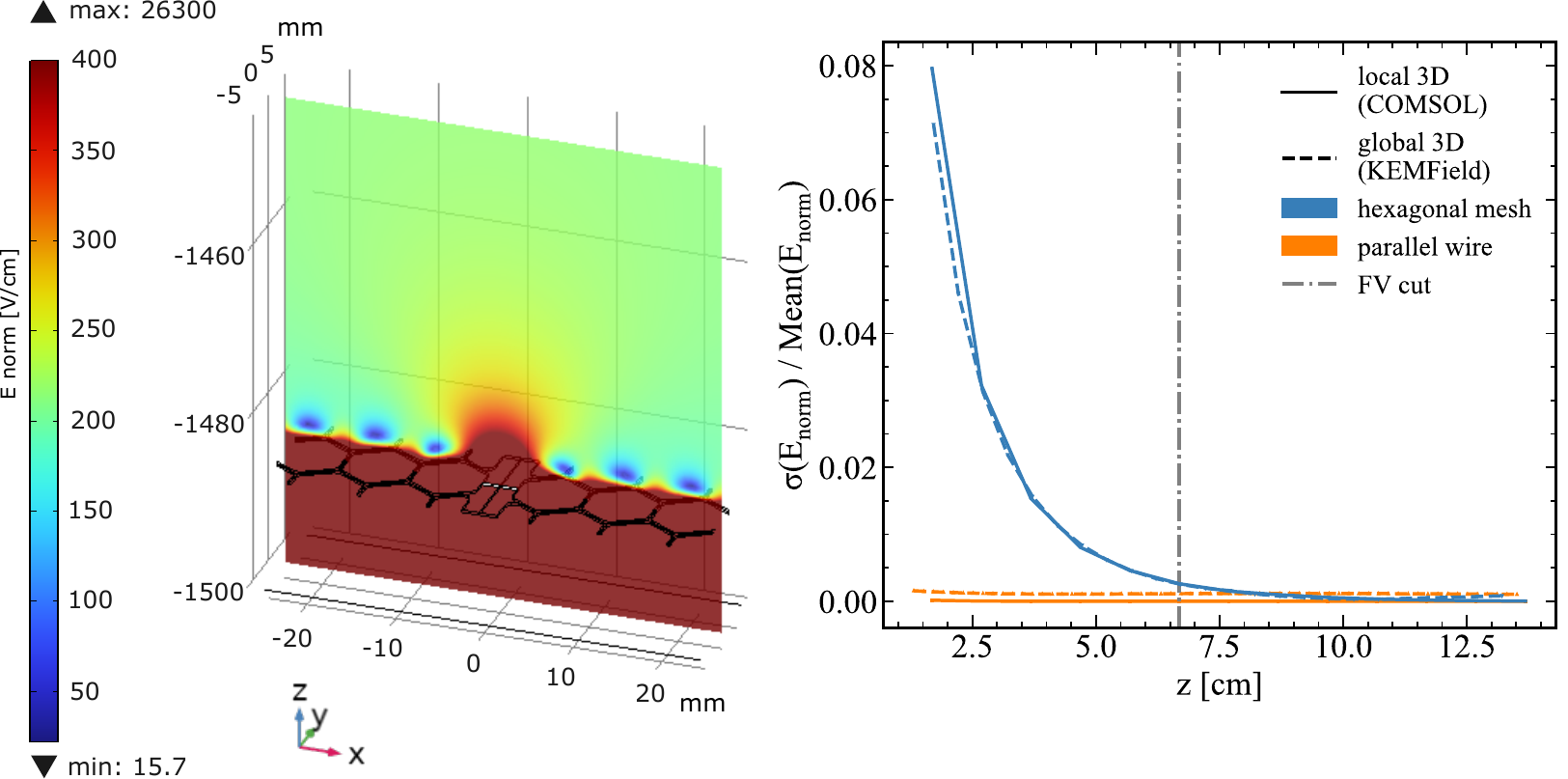}
\caption{Left: close-up of the result of the local 3D simulation with COMSOL\textsuperscript{\textregistered}. The welded strip is along the y-direction. The $E_{norm}$ field on the x-z plane at $y=0$ is also shown by the colour map. Right: the standard deviation of $E_{norm}$ for $x\in[-40, 40]$\SI{}{\,mm} normalized by the mean field at the same z-distance from the cathode. 
The field inhomogeneity due to the welded strip becomes negligible at 7.5\,cm above the cathode.
Results from both the local 3D (solid line) and global 3D (dashed line) simulations from COMSOL\textsuperscript{\textregistered} and KEMField are shown. The color indicates the electrode configuration, with a hexagonal mesh cathode in blue and a parallel wire cathode in orange.
The fiducial volume cut used in a previous XENONnT analysis \cite{nT_analysis_SR0:2025} is also indicated by the dashed-dotted line.}
\label{fig:welding_seam_COMSOL}
\end{figure}

Subsequently, a local 3D simulation with finer discretization and accurate geometry was performed using the Electrostatics module of COMSOL Multiphysics\textsuperscript{\textregistered} software \cite{COMSOL}. 
The simulation domain is shown in \autoref{fig:welding_seam_COMSOL}~(left). 
The hexagonal mesh geometry was imported from the CAD drawing.
The top and bottom boundary conditions were retrieved from the aforementioned full 3D field simulation.
Zero-charged boundary conditions were imposed on the four sides, meaning the simulation domain repeated infinitely. 
\begin{table}[htb!]
\caption{Requirements for the mesh electrode as a cathode.
The maximum sagging of the mesh under electrostatic and gravitational force ($h_{\text{max}}$) was estimated based on the electrode configuration in the XENONnT detector. 
The yield tensile strength for the mesh ($\text{YTS}_{0.2\%}$) was taken as the typical value of SS304 from \cite{SS_book}. 
The thermal stress ($\sigma_s$) is defined as in~\autoref{tab:anode_requirement}.
The stress due to the tension applied to the mesh is defined as $\sigma_T$. } \label{tab:cathode_requirement}
\begin{center}
\begin{tabular}{l l l}
Quantity & Symbol & Requirement\\
\hline
Sagging limit & $h_{\text{max}} $& $<$\,\SI{2.0}{mm} \\
Safety factor & $\text{YTS}_{0.2\%}\,/\,(\sigma_T + \sigma_s)$& $ \gtrsim 1.5$ \\
Transparency & & $\gtrsim 90\%$\\
\end{tabular}
\end{center}
\end{table}

To evaluate the result in both simulations, we consider an array of lines perpendicular to the welded strip at $y=0$ and $x\in[-40,40]$\SI{}{mm}, parallel in z-direction.
The normalized standard deviation of the field magnitude along these lines was used as the metric for field homogeneity. 
In comparison, the same calculation was performed for parallel wires without the central welded strip. 
The result is shown in \autoref{fig:welding_seam_COMSOL}.
Compared to the parallel wire case, the case with the welded strip has a larger deviation in the field as expected. 
The closer it is to the cathode electrode, the larger the field inhomogeneity. 
However, the break even point is around 7.5 cm above the cathode, and is comparable to the previous XENONnT fiducial volume bottom boundary \cite{nT_analysis_SR0:2025}.

The electrostatic force on the cathode was derived using a similar COMSOL local 3D simulation with an infinitely large cathode, but without the welded strip in place. 
The resulting load was \SI{1.5}{N/m^2}. 
Additional loads, namely tensioning stress, gravitational loads, and thermal shrinkage, were also considered for the mechanical simulation.
Similar to \autoref{sec:mech_sim_frame}, the ANSYS Workbench 2021 was used to determine the stress and the safety factor of the hexagonal mesh under loads.
To reduce the computational time, a 2-fold mirror symmetry and a scale factor of 2 were applied to the model. 
With a simpler model, we verified that the scale factor did not change the result as long as the mass of the electrode remained unchanged. 

From the result, the maximum stress ($\sigma_T + \sigma_s$) is at \SI{167}{MPa} with a \SI{25}{K} temperature difference between the mesh and the frame. 
This corresponds to a safety factor of 1.4, assuming the YTS for the material to be at \SI{240}{MPa} \cite{SS_book}. 
The expected sagging of the mesh during operation is well below the maximally allowed sagging of 2\,mm as shown in \autoref{sec:hexmesh_assembly}.

Comparing the mentioned hexagonal mesh of the leg width of \SI{0.3}{mm}, an opening of \SI{7.5}{mm} and a \SI{4}{mm}-welded strip at the center, to a parallel-wire electrode with \SI{0.3}{mm} diameter wires with the pitch of \SI{7.5}{mm}, the transparency is lowered by 4\% reaching $91.4\%$.
This can be used as a proxy for the first-order effect on the Light Collection Efficiency (LCE): the probability of detecting a photon produced in the TPC. 

\par
A second-order effect is the impact of the shadow cast by the electrode onto the PMT array.
To study this electrode shadow, we set up a Monte-Carlo simulation of a complete XLZD-scale TPC using GEANT4 \cite{geant4}. 
We implemented four different electrode designs: An opaque disk with transparency equal to the transparent area of a typical electrode, an electrode made of parallel wires, one made of parallel and perpendicular wires forming a square pattern, and an electrode with a hexagonal mesh pattern.
\par
\begin{figure}[htb]
\centering
\includegraphics[width=.9\textwidth]{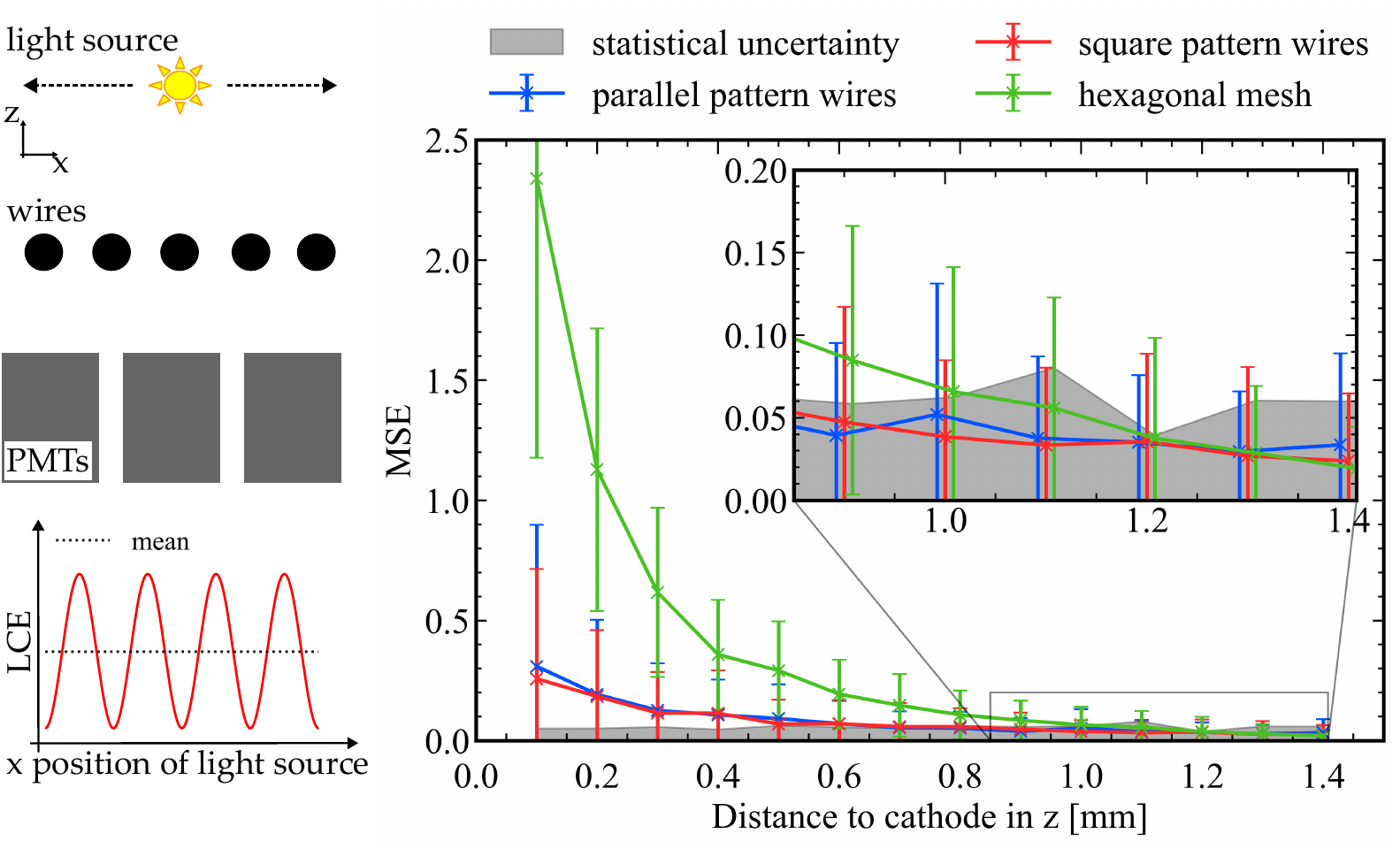}
\caption{Left: Sketch of the quantification of the shadowing effect. The light collection efficiency (LCE), here visualized as a generic periodic function, varies more as a function of $x$ if the light source is closer to the wires in $z$. Right: Impact of the electrode shadow in terms of the Mean Square Error (MSE) of the LCE as a function of the distance to the electrode. 
The error bars are the standard deviation of the square error.
The gray band shows the MSE of an opaque disc approximation, where deviations from the mean are caused only by statistical fluctuations.
}
\label{fig:shadowing-mse}
\end{figure}

We have then simulated a light source in a line parallel ($\vec{x}$-direction) to the electrode at multiple distances from the electrode ($\vec{z}$-direction).
For each $\vec{z}$ distance from the electrode we calculated the Mean Square Error (MSE) of the LCE:
\begin{equation}
    \text{MSE}(\vec{z}) = \langle (\text{LCE}(\vec{x},\vec{z}) - \hat{\text{LCE}}(\vec{z}))^{2} \rangle_{\vec{x}}
\end{equation}
The MSE is high if along $\vec{x}$ there is a large deviation in the LCE and therefore serves as a measure of the impact of the electrode shadow.
The opaque disk design serves as a baseline, since it casts no geometric shadow and therefore all variations in LCE are solely caused by statistical uncertainty.
The MSE as a function of $\vec{z}$ distance to the cathode is showcased in \autoref{fig:shadowing-mse}.
For distances greater than $\SI{1}{mm}$ from the electrode, the shadowing effect is less than the statistical variation of the LCE. 
This result shows that the electrode shadow has a negligible effect on the LCE and does not need to be taken into account during optimization of future electrode geometries.

\subsection{Tensioning \& Assembly Mechanism } \label{sec:hexmesh_assembly}

\begin{figure}[htb]
\centering
\includegraphics[width=\textwidth]{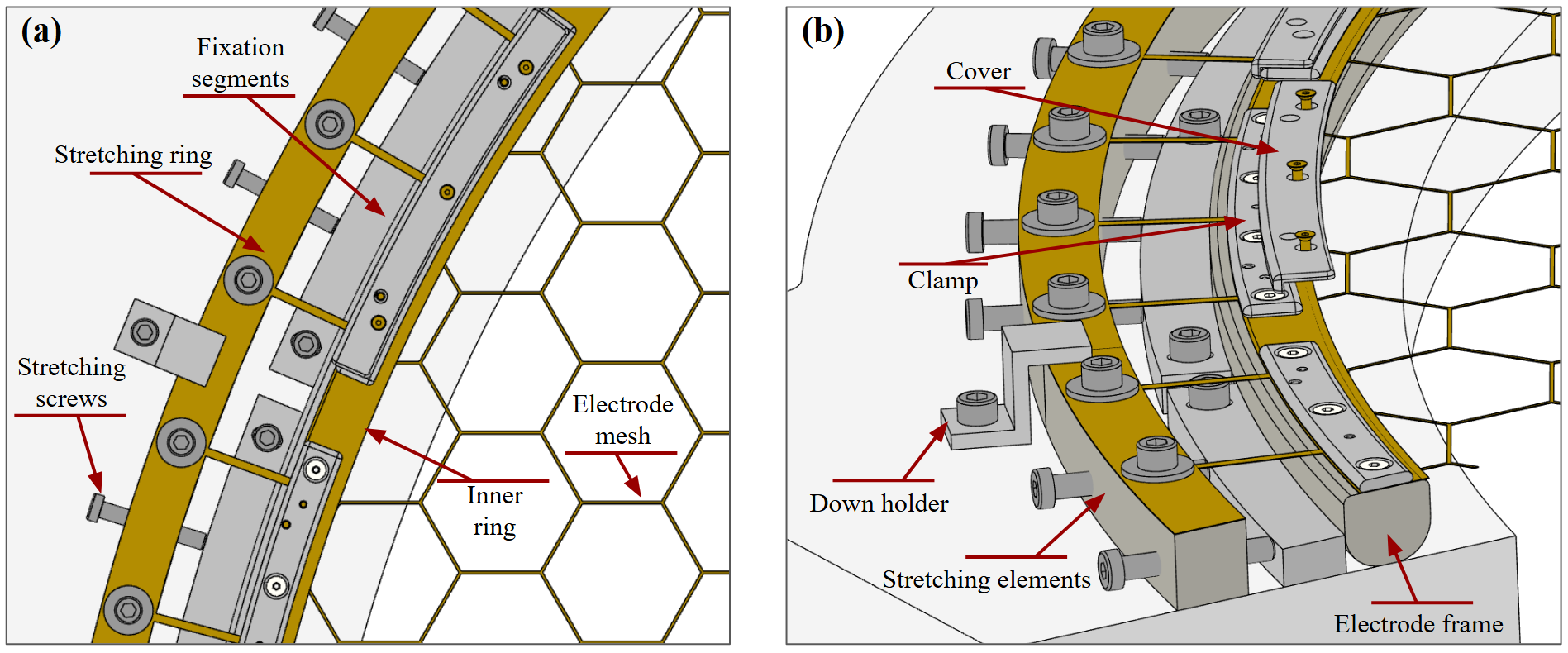}
\caption{Top and cutaway view of the stretching and fixation mechanism of the electrode mesh onto an electrode frame. 
The electrode frame is shown on top of a dedicated SS plate (light gray). 
Note that the displayed hexagonal mesh segments are not to scale and are shown for illustration purposes only.
In image (a) a top view of the stretching mechanism is shown, while image (b) depicts  cross-section view of the stretching mechanism, showcasing the SS covers fixed with M4 Ag-coated screws (white-colored screws) and clamping pieces fixed with M2.5 Au-coated screws (gold-colored screws).
}
\label{fig:mesh_assembly}
\end{figure}

A dedicated procedure and tools were developed for mounting, stretching and fixation of the hexagonal mesh onto the electrode frame.
The electrode frame was placed on a custom designed SS mounting plate with dedicated threaded holes at regular intervals for hosting the 12 Z-shaped down holder elements, as depicted in \autoref{fig:mesh_assembly}.
Along the outer radius of the electrode frame 12 fixation segments were mounted on the plate preventing the electrode frame from moving during the stretching process.
Subsequently, the stretching ring of the electrode mesh was attached to the 12 stretching elements via screws.
The stretching screws were then gradually turned pushing against the fixation segments in a controlled fashion, stretching the mesh.
The sole role of the down holders along the radius of the assembly setup was to prevent the perpendicular movement of the stretching elements.

After the stretching process was complete, the mesh was fixed to the electrode frame using dedicated 24 clamping pieces, and vented M4 Ag-coated screws.
Then, all the ancillary segments were removed and the stretching ring was cut off by severing the connections between the stretching ring and mesh's inner ring. 
To conceal the cut area and the clamps, additional 24 cover pieces were placed on top of each clamp and fixed with vented M2.5 Au-coated screws.  
The assembly process was first tested at KIT in non-cleanroom conditions, and then was carried out inside a cleanroom at LNGS.
The sagging of the mesh throughout the stretching process was verified by measuring the distance between the mesh's surface and a reference surface at the center of the SS mounting plate.
The sagging of the mesh was controlled at every stage of the installation until mounting the electrode in the detector. 
The measured sagging of 0.6\,mm is expected to lead to a maximum sagging $h_{\text{max}}$ of 1.1\,mm during operation, well below the requirement of 2\,mm, as given in \autoref{tab:cathode_requirement}.

\subsection{ML-based Defect Detection} \label{sec:ML_defect}

\begin{figure}[htb]
\centering
\includegraphics[width=1\textwidth]{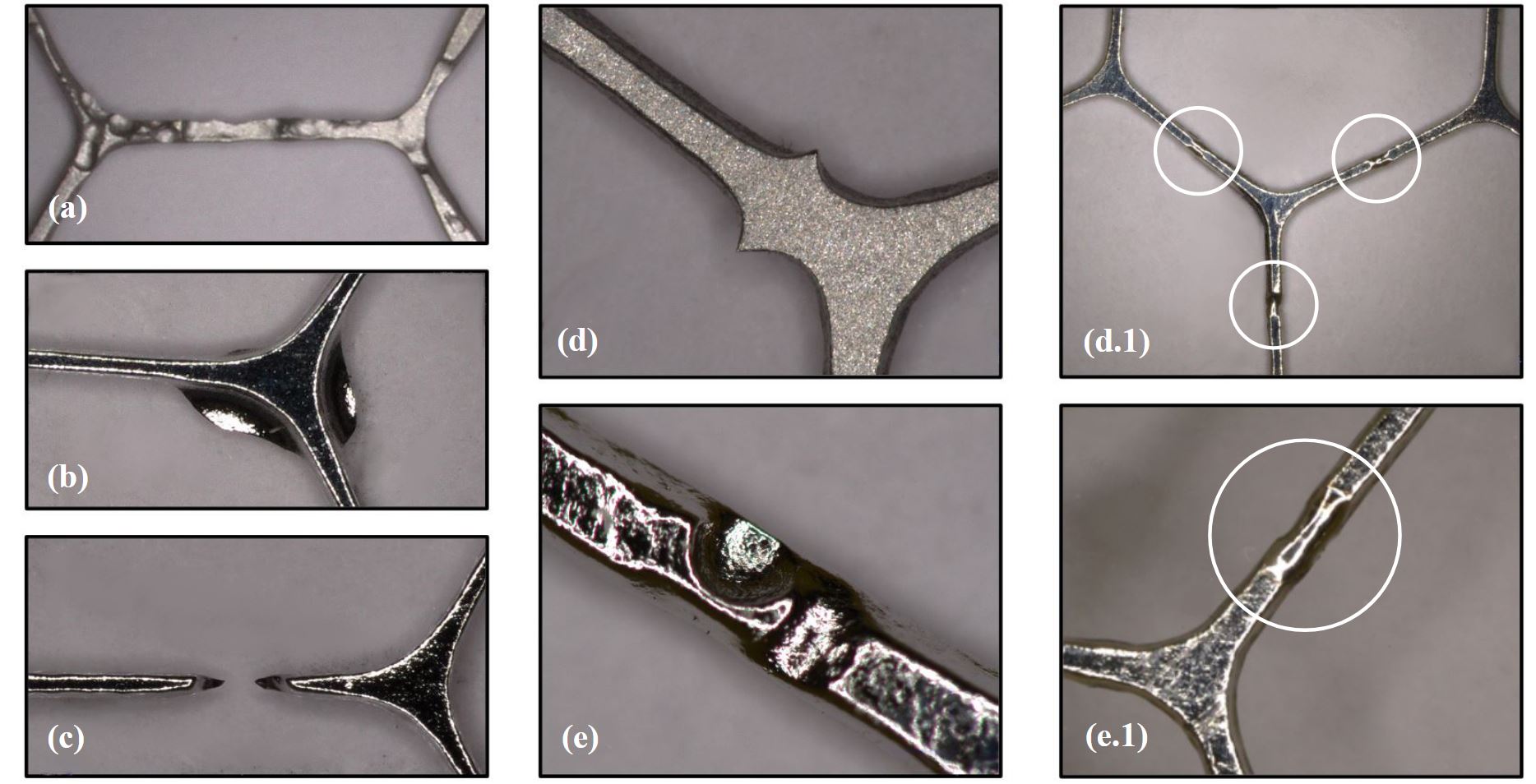}
\caption{
Examples for different types of features that were encountered on the hex meshes during inspection. 
The displayed features were all induced during the production of the electrode mesh.
Features shown in panels (a) and (e) constitute missing material, while features of type shown in panel (b) are representative of added material.
Several features exhibited sharp corners, such as the cut in the mesh shown in panel (c) and the spike shown in panel (d).
Panels (d.1) and (e.1) showcase the features shown in panels (d) and (e) after the treatment and repair procedure, respectively. 
In the case of feature shown in panel (d) its location was fully replaced by a donor piece from another electropolished mesh, while the feature shown in panel (e) was laser welded in place.
In both cases the laser welding locations are highlighted with white circles.
Features in panel (b) and (c) were found on a spare electrode mesh, and are showcased here for illustration purposes.
For visualization purposes the images in all panels were digitally processed, cleaning the gray background in each image.}
\label{fig:features}
\end{figure}

During production or handling, a mesh can be damaged, potentially deteriorating the mechanical stability of the mesh, or locally altering the produced electric field eventually leading to electron emission.
To be able to assess the impact of different features, such as the ones seen in \autoref{fig:features}, they must first be identified.
A typical mesh for a large-scale electrode can consist of hundreds of thousands of individual legs, making a full inspection by eye unfeasible.
Reconstruction based anomaly detection using Machine Learning (ML) models can be used to find physical defects in images of workpieces \cite{yolov5-defect-detection,cascading-ae-defect,granite_optical}.

\begin{figure}[htb]
\centering
\includegraphics[width=\textwidth]{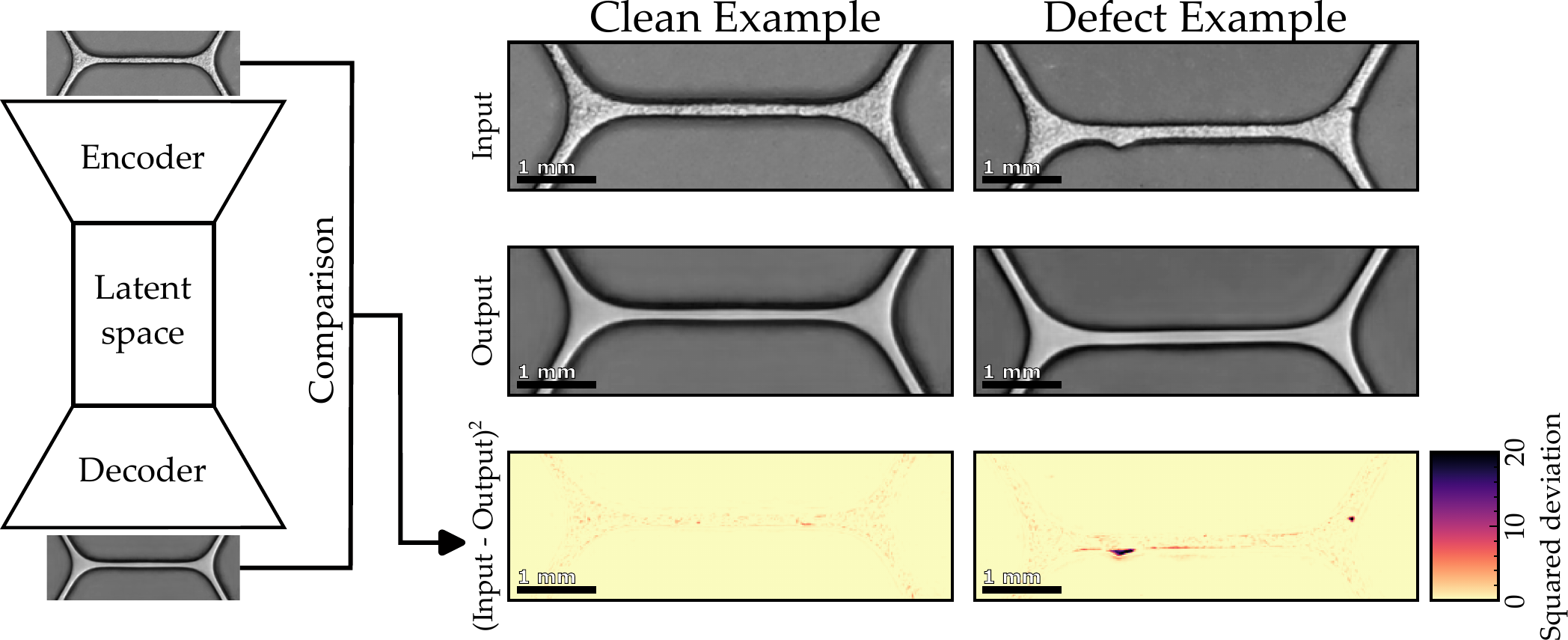}
\caption{Left: A sketch of the working principle. 
The ML model encodes the input image into the lower dimensional latent space via an encoder network and then constructs a resemblance of the initial input via a decoder network. 
Anomalies too rare to be learned by either network show up as deviations between input and output. 
Right: Two examples of input, output and squared deviation, one for a clean and one for a defect example.
The length scale indicating one millimeter was added to the images post-hoc and not included when propagating the images through the ML pipeline.
The deviation between input and output of the clean example is low across all pixels. 
In the defect example the spike structure on the left side of the leg, and the notch structure on the right side of the leg are clearly visible as regions of high squared deviation.}
\label{fig:ml-defect-example}
\end{figure}

We used a Variational Autoencoder (VAE) \cite{vae-introduction} model, a probabilistic extension of the autoencoder (AE) \cite{ae-introduction} architecture.
An AE consists of two networks: An encoder and a decoder.
The encoder transforms a high-dimensional input $x\in\mathbb{R}^{n}$ into a lower-dimensional set of variables, called the latent space $z\in\mathbb{R}^{m}$ ($m\ll n$). 
Afterwards, the decoder transforms the latent space into a reconstruction $\tilde{x}\in\mathbb{R}^{n}$ of the same shape as the input.
In our case, the input and output were images of individual legs of dimension $n=\qtyproduct{300x100}{px}$.
The underlying assumption was that features in the mesh were too rare for the encoder to successfully learn to encode them in $z$, and therefore could not be reconstructed by the decoder.
Given an input featuring some anomaly, the AE reconstructed a clean leg, allowing us to detect the anomaly by inspecting the squared difference between input and output $(x - \tilde{x})^{2}$.
This working principle, including a clean example and an example showing a small feature, is visualized in \autoref{fig:ml-defect-example}.

\begin{figure}[htb]
\centering
\includegraphics[width=\textwidth]{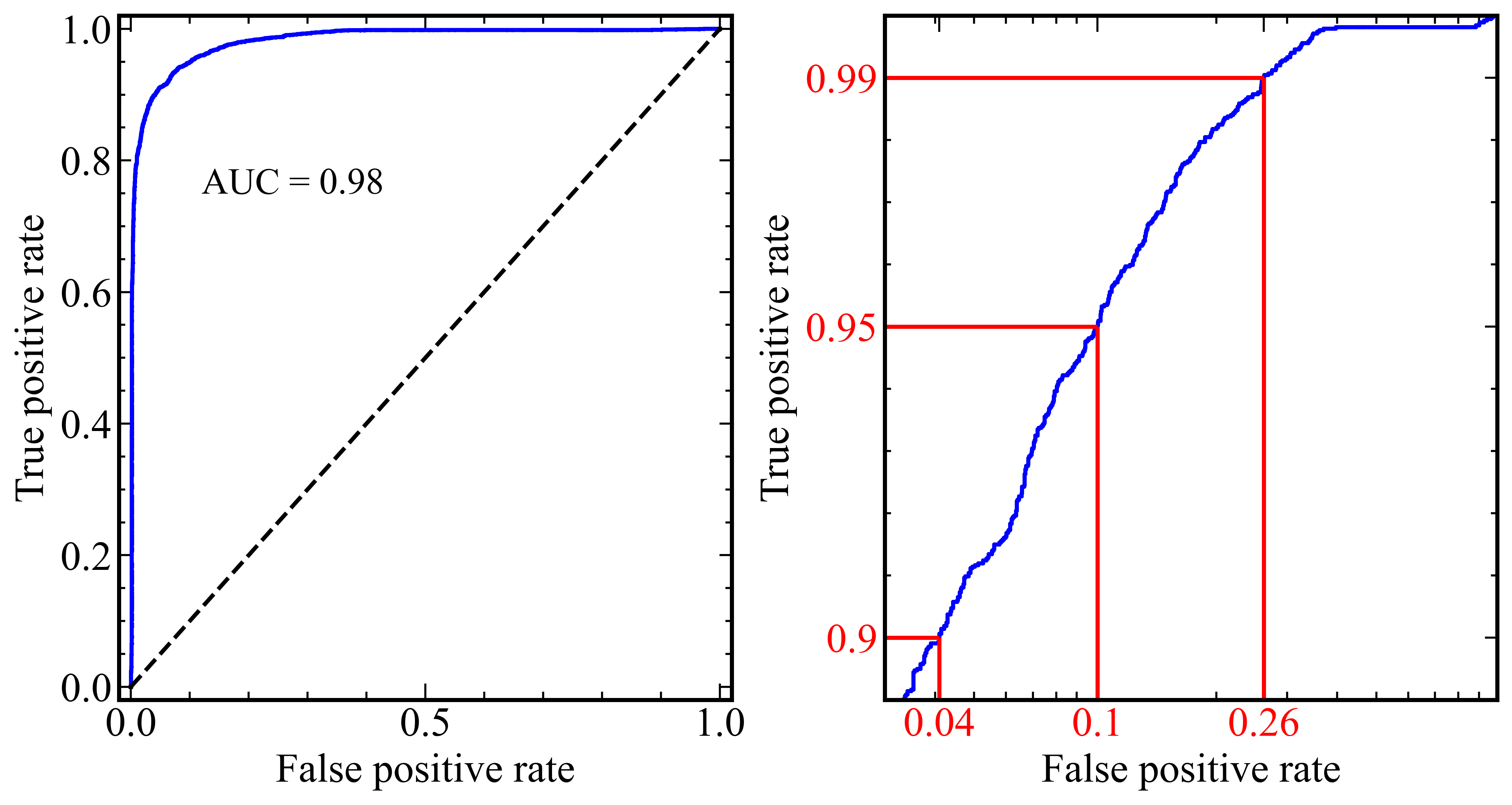}
\caption{ROC curve of the defect detection model (blue). 
The true positive rate corresponds to the efficiency of the model to detect a defect.
The false positive rate is the probability of the model to declare a clean leg as defect.
The black dashed line is the theoretical performance of a randomly guessing estimator.
The area under the ROC curve (AUC) is 0.98. 
The right panel is zoomed-in on the region of high defect detection efficiency. 
The red lines signify examples of desired efficiencies (vertical axis) and their corresponding false positive rates (horizontal axis). 
Note that the horizontal axis in the right panel is in log-scale.}
\label{fig:defect-roc-curve}
\end{figure}

In a VAE, the latent variable $z$ is replaced by $m$ uncorrelated Normal Distributions, parametrized by a mean value $\vec{\mu}\in\mathbb{R}^{m}$ and a standard deviation $\vec{\sigma}\in\mathbb{R}_{+}^{m}$, which are learned by the encoder.
The decoder then samples from the input-dependent multivariate Normal Distribution $z\sim\mathcal{N}(\vec{\mu},\vec{\sigma}|x)$. 
The VAE in this work was trained on the Evidence Lower Bound loss function, as is usual for such models \cite{vae-introduction}.
The performance of the VAE was evaluated on 4000 clean legs and 4000 digitally augmented images of defect legs.
\autoref{fig:defect-roc-curve} shows the Receiver-Operating-Characteristic (ROC) of the VAE model, where the positive case corresponds to an image being identified as showing a feature.
The Area Under the Curve of the VAEs ROC is $\text{AUC} = 0.98$, which is comparable to other defect detection models \cite{mem-ae}.

While this model offers certain advantages, there is no confirmation of whether or not all features were detected, and if a detected feature is actually detrimental to the performance of the mesh. 
To quantify the effect of identified features additional performance tests under high voltage conditions were performed, shown in \autoref{sec:HV_test}.

\subsection{Feature Treatment \& Repair } \label{sec:defects_features}

Features identified on the surface of the mesh were treated and repaired in several ways.
Some of the features were mechanically ground down and polished, reducing their sharpness and the potential danger of field emission or breakdown.
Additionally, several mesh segments with features, such as the one illustrated in \autoref{fig:features} (a) and (d),  could not be ground down and were fully replaced.
To that end, mesh segments with such features were carefully cut from the body of the mesh.
Subsequently, as shown in \autoref{fig:features} (d.1), donor segments which were cut from a spare mesh were laser welded in place using the same laser welding technique which was used to weld the half meshes together.
Furthermore, for features such as the one seen in \autoref{fig:features} (e) laser welding was used to fill the location resulting in a connection seen in \autoref{fig:features} (e.1).

Features whose shape or size was not deemed as potentially detrimental to the mesh's future performance under high-voltage were not treated or repaired.
Prior to electropolishing, 7 features were identified out of which 6 were treated or repaired and the mesh was subsequently electropolished.
Throughout the rest of the text this is referred to as stage 1 repair.
After electropolishing, through manual inspection and with the use of ML techniques, 6 more features were identified, out of which 4 were treated.
This is referred to as stage 2 repair, after which the mesh was electropolished again.

Robustness of the laser welded connections was verified by performing stress tests.
Non electropolished mesh segments were laser welded together and pulled in opposite directions while recording the force until a fracture occurred.
For laser welded segments the fracture happened at the welded connection, and the estimated UTS decreased from \SI{600}{MPa} to \SI{454}{MPa} when compared to non-welded mesh segments. 
Nonetheless, the deduced UTS values lie well beyond the load on the electrode that was estimated in \autoref{sec:hex_mesh_sims} to be \SI{167}{MPa}.

\section{High-Voltage Testing \& Performance}
\label{sec:HV_test}

Features, such as the ones showcased in \autoref{sec:ML_defect}, could lead to localized field enhancement, which in turn could result in detrimental effects such as field emission and even breakdown \cite{adam_bailey_thesis_em, pixie}.
To assess the performance of an electrode, one cannot rely solely on electrostatic simulations, as the field around each feature is highly dependent on its geometry.
Hence, to study such effects and to determine the overall performance of the electrode, we have performed a series of HV tests in a gaseous argon (GAr) environment.
In those tests, we aimed to observe light associated with effects such as field emission or breakdown events.

The tests were conducted in a light-tight Faraday cage and utilized a FuG Elektronik HCP 140 power supply, which provides HV of up to \SI{200}{kV} DC (negative polarity) with a maximum current of \SI{0.7}{mA}.
The HV is fed from the power supply into the Faraday cage via a shielded cable, which connects to a custom-developed ground terminator with an aluminum toroid at its top. 
A LabVIEW-based software facilitates control and data readout from the power supply at frequencies of the order of \SI{1}{Hz}.
Camera imaging and data from the power supply were used to identify areas associated with potential field emission and to assess the reliability of the electrode.

\subsection{High-Voltage Test Setup }
\label{sec:hv_test_setup}

\begin{figure}[htb] 
\centering
\includegraphics[width=\textwidth]{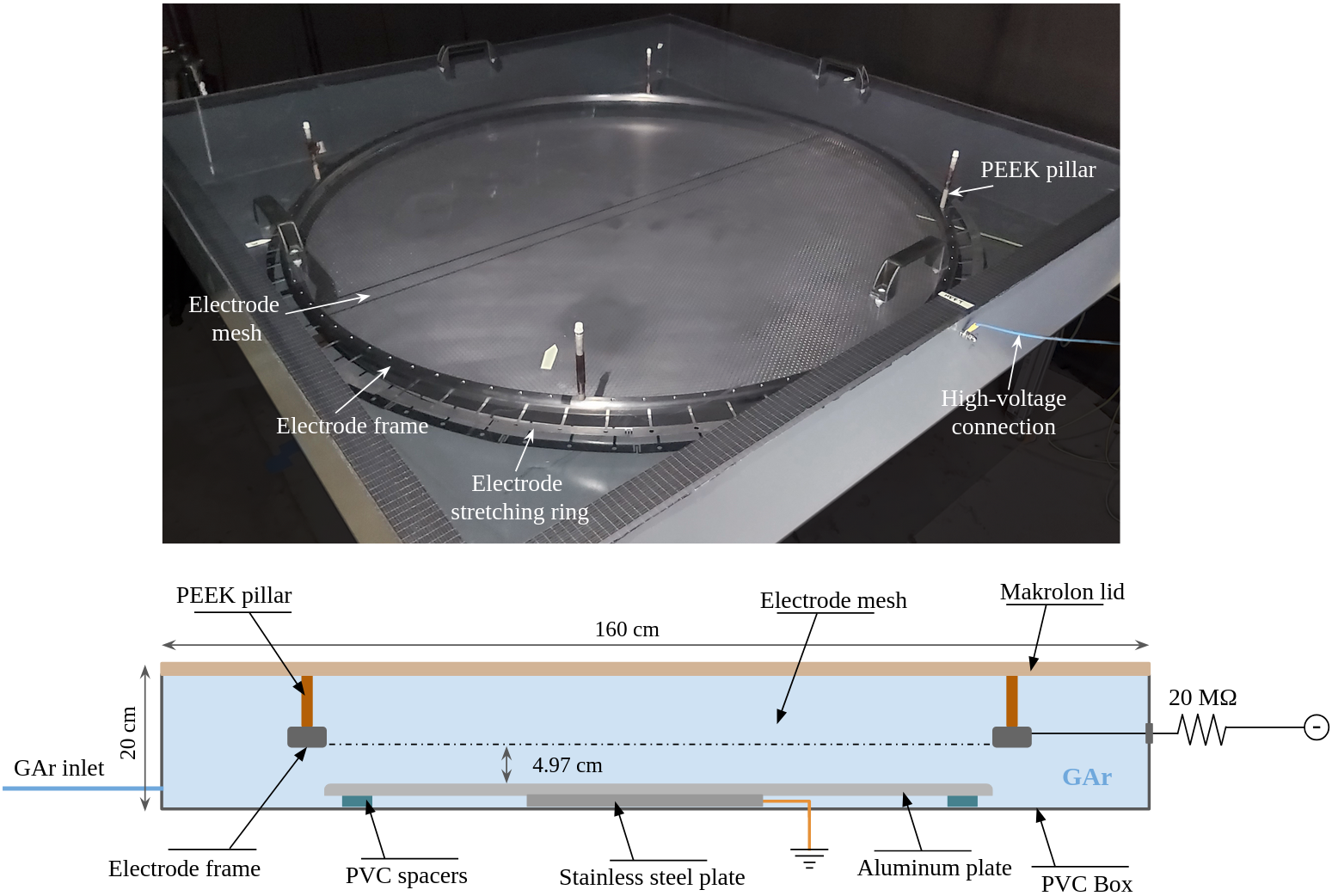}
\caption{Final configuration of the HV test setup for evaluation of electrode performance in GAr.
Top: A photograph of the PVC test box within the light-tight Faraday cage. 
In this configuration the electrode frame was suspended inside the box from four PEEK pillars. 
To prevent current leakage on the surface of the pillars their centers were also wrapped in Kapton tape. 
Imperfections on the surface of the aluminum ground plate can be seen as stain-like patterns on its surface.
Bottom: A schematic diagram of the test box HV setup (not to scale). 
To limit the power input into the system and prevent damage to the tested electrode mesh, the HV was passed to the electrode under test via 2 resistors with \SI{10}{M\Omega} each.}
\label{fig:HVtest_setup}
\end{figure}

A test setup, inspired by the one described in \cite{daniel_wenz_thesis,Mitra2023}, was built inside the Faraday cage.
As illustrated in \autoref{fig:HVtest_setup}, at the heart of the setup lies a square PVC box measuring \SI{1.6}{m} in length and \SI{20}{cm} in depth. 
The box was elevated above the floor with the use of a dedicated four-legged frame with wheels.
To maintain a GAr atmosphere and protect the electrode under test from environmental contaminants the box was instrumented with a \SI{1.1}{cm} thick lid made of Makroclear\textsuperscript{\textregistered} polycarbonate sheet \cite{makroclear}, with strips of Armaflex\textsuperscript{\textregistered} tape for sealing.
The use of the polycarbonate lid transmits the light in the visible and in the near-infrared range \cite{makroclear_transmittance, polycarbonate_transmittance:2018}, allowing us to observe light, which could be associated with effects such as neutral bremsstrahlung electroluminescence and the avalanche scintillation of GAr in the near-infrared range \cite{ELreview2020, gAr_spectrum:2022}.
The box was fitted with a GAr tube inlet, a HV connection to the electrode and a ground connection.
The flow of GAr into the box was controlled via a flow meter connected to the bottles containing grade 5.0 argon.
Additionally, an Infactory NC-7004-675 humidity and temperature sensor \cite{infactory} was placed at the farthest corner from the GAr inlet, as these parameters can affect the breakdown voltage. 
Since the PVC box was not leak-tight, the pressure inside the box was assumed to be close to the atmospheric pressure, and thus not monitored. 

A \SI{1.2}{m} in diameter and \SI{5}{mm} thick matt aluminum plate with a rounded edge was placed at the bottom of the box on top of a grounded SS plate, forming a ground plate in the setup.
A number of PVC spacers was used to further support the aluminum plate on its sides, resulting in a total estimated elevation of \SI{1}{cm} of the plate above the box's bottom.
The maximum measured gap distance in between the electrode under test and the ground plate was \SI{4.97}{cm}.
This measured gap distance was used to define the so-called bulk field in subsequent analysis.
However, it should be noted that throughout the HV tests the precise distance and alignment between the ground plate and the tested electrode were not controlled. 
The unevenness of the ground plate, defined as the vertical distance between its highest and lowest points, was measured to be \SI{6.4}{mm}, amounting to about 15\% of the gap distance.

The electrode mesh was mounted on the electrode frame in non-cleanroom environment conditions, with the use of the procedure described in \autoref{sec:hexmesh_assembly}.
However, as depicted in \autoref{fig:HVtest_setup}~(top), for the purpose of the HV tests the stretching ring was not removed and the SS covers were not installed.
The electrode was positioned in the setup with the side on which the mesh was mounted facing down towards the ground plate.
To facilitate the HV connection, the electrode was connected to a connection in the box's wall, which subsequently was connected to the toroid.
For most of the tests, \SI{20}{\Mohm} resistors were added on the HV line leading from the test box to the toroid, limiting the current into the system and preventing potential electrode aging effects as a result of discharges \cite{electrode_againg_1973}. 
In the final test configuration, the electrode frame was suspended from the box's lid using PEEK \cite{insulator:2018} pillars wrapped in Kapton tape, as illustrated in \autoref{fig:HVtest_setup}.
Prior to that, several insulators with a varying geometry, produced from PVC, POM, and 3D-printed ASA plastic were tested as separators between the electrode frame and the ground plate.
However, their use resulted in frequent breakdowns at the insulators, indicating that these insulators enhanced locally the field along the edge of the ground plate. 

Two cameras were installed above the test box, positioned roughly at its center, such that the field of view covered the entire electrode.\footnote{A Nikon Z\,50 mirrorless camera, a Canon EOS 90D, and a Canon EOS\,700D in rotation, with the NIKKOR\,DX 18-140mm f/3.5-6.3, Canon EF-S 18-135mm f/3.5-5.6 IS USM, and Canon EF-S 18-135mm f/3.5-5.6 IS STM lenses, respectively.} 
To prioritize capturing the HV-induced dim light most images were taken with a long-exposure of 5-10\,s. 
In several tests a third camera was introduced to capture videos which provide better time resolution, albeit with reduced exposure and thus light collection.

\begin{table}[hbt]
\caption{Tests that were included in the analysis, with corresponding GAr flush time at a rate of 20\,NL/min before the test, and hardware variation (insulator type, external resistor) that were used in each test. Stages 1 and 2 are defined in \autoref{sec:defects_features}.
}
\label{tab:run_list}
\begin{center}
\begin{tabular}{ l|l|l l l l l }
        &       & No. of  &                 & External & GAr Flush \\
Test ID & Stage & Ramps   & Insulator Type  & Resistor &  Time [\SI{}{\hour}] \\\hline
\hline
1   &  1 &  4   & 3D-printed spacer   & No        & 2.7   \\ \hline
2   &  1 &  2   & PEEK pillars        & Yes    & 1.5   \\ \hline
3   &  1 &  11  & PEEK pillars        & Yes       & 1.9   \\ \hline
4   &  2 &  5   & PEEK pillars        & Yes       & 2.0   \\
\end{tabular}
\end{center}

\end{table}

The HV tests were performed in the following manner.
Before each HV test, we continuously flushed the closed PVC box with GAr at a rate of 20\,NL/min.
Due to the high Reynolds number of around \SI{5e5}{}, the gas mixture was considered to be reasonably well mixed, and no further mechanism was used to mix the gas inside the box. 
Then, the aforementioned cameras were installed and calibrated as described in \autoref{sec:calibration}.
Each HV test consisted of multiple ramps. 
During each ramp the biasing voltage of the electrode was increased in steps while gradually reducing the ramping speed at higher voltages.
The current limit was set at below \SI{5}{\micro A}.
At the same time, the current and voltage were tracked via the software system controlling the power supply, and the situation inside the box was monitored via a live feed from the cameras.
In the event of a breakdown, or persistent current fluctuations, which were considered as an indication of a potential breakdown \cite{xebra2023}, we manually ramped down the HV.
The conducted tests, their association with the electrode mesh treatment stages, as well as their constituent ramps and a selection of key test parameters are listed in \autoref{tab:run_list}.

\subsection{Calibration \& Tests} \label{sec:calibration}

\subsubsection{Camera Calibration}
We calibrated the time, noise levels, and the position of the electrode on each camera. 
Since the internal clocks of the cameras were not synchronized, the time difference between the internal clocks and the GPS time was recorded before each HV test. 
For noise calibration, images were taken with the camera's lens cap on and the cameras placed in a dark environment. 
Pixels that repeatedly showed a signal in these dark calibration images were deemed noisy and excluded from the analysis on the individual measurement day.
Position calibration images were taken while the electrode was illuminated, allowing us to subsequently correlate the location of the glow or breakdown in the camera imagery to their position on the tested electrode. 

\subsubsection{Argon Calibration}
Breakdown voltage or HV instability depends, among other factors, on the gas composition of the environment as well as the water content in it \cite{humidity, humidity_2, high_voltage_book:1991}.
Therefore, the humidity inside the test box was monitored by a humidity sensor. 
It was also used as a proxy for the air concentration as the GAr was flushed into the box.
Note that the breakdown voltage in argon gas is lower than that in air at our operating conditions \cite{Raizer_book:1991}; therefore, the humidity should correlate with the breakdown voltage in our setup. 
To that end, a calibration run was performed to verify the correlation between GAr flush time, humidity, and the HV instability of the setup. 

\begin{figure}[htb]
\centering
\includegraphics[width=\textwidth]{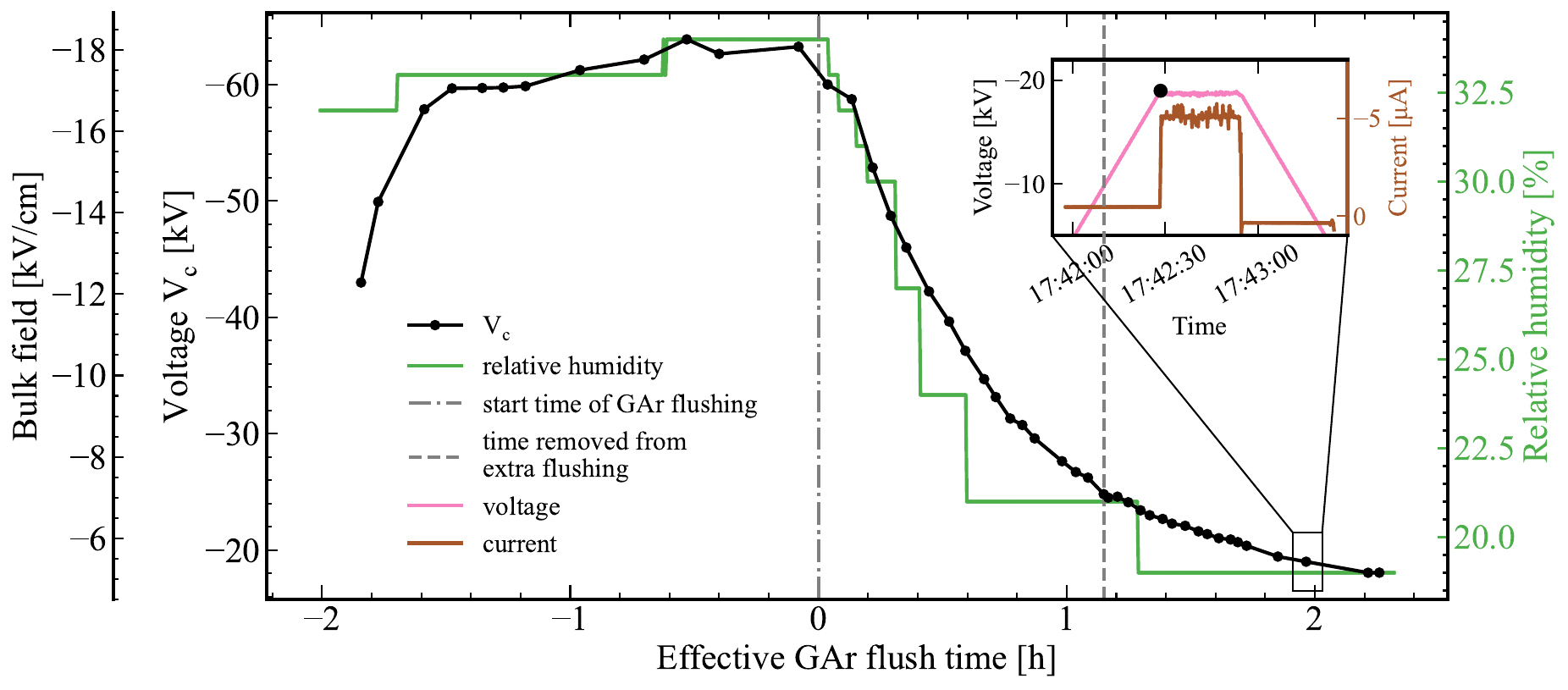}
\caption{Evolution of the critical voltage $V_c$ with respect to the effective GAr flush time. The inset shows an example for one ramp of $V_c$, which is the voltage right before the current reached \SI{5}{\micro\ampere}. The y-axis (inner left) shows the voltage, $V_c$. The y-axis (outer left) shows the reference bulk field, deduced as $V_c$ divided by the electrode gap. The x-axis is the effective elapsed time of the GAr flush. At time $t=0$ (marked by the dash-dotted line), the GAr was flushing at 20\,NL/min. After around one hour of flushing (marked by the dashed line), there was a 20-hour pause on flushing, where some GAr escaped. The additional flush time needed to reduce $V_c$ again was removed from the above plot. The right y-axis indicates the humidity level measured by the sensor at the synchronized time. }
\label{fig:argon_time}
\end{figure}

The calibration setup consists of one biased and one grounded SS plate, with outer diameters of \SI{29.5}{cm} and \SI{16.5}{cm}, respectively. 
The HV was connected to the setup via the \SI{20}{\Mohm} resistor as in the actual tests.
The biased plate was supported by three PEEK pillars mounted on a metal plate of \SI{10.5}{cm}.
The gap distance was measured to be $35 \pm 1$\,\SI{}{mm} without controlling the alignment.
The setup was assembled in a non-cleanroom environment but wiped with ethanol and subsequently positioned at the farthest corner away from the gas inlet. 
A humidity sensor was placed about 10\,cm from the biased plate, further away from the gas inlet, to measure the humidity inside the box. 
During calibration, while argon was flushed into the box at 20\,NL/min, we repeatedly ramped up the HV until the current limit at \SI{5}{\micro\ampere} was reached.  
The critical voltage $V_c$ is defined as the voltage one time-step before the current limit was reached, as shown in the inset plot in \autoref{fig:argon_time}. 
$V_c$ is plotted as a function of the effective GAr flush time, as shown in the same figure. 

In this calibration run, most ramps have a ramp-up speed of 500 V/s. 
At the beginning of the calibration campaign (negative times in \autoref{fig:argon_time}), many ramps were performed to condition the system, removing the remaining dust and impurities \cite{lucie:2019, conditioning:1982}, resulting in a gradual increase in $V_c$.
At time $=0$, the GAr valve was opened and both the humidity and $V_c$ were reduced.
Since there was an overnight break during the test, the argon escaped from the box, causing $V_c$ to increase. Therefore, an extra flush time is required to bring $V_c$ back to the level it was before the pause.
The dashed line in \autoref{fig:argon_time} marks the time window that was eliminated.
The argon flow was closed occasionally to confirm that the change in GAr concentration caused the reduction in $V_c$.
Those periods were also removed from the plot such that we obtained the effective flush time.

The $V_c$ and humidity profiles were correlated throughout the test, the latter being limited by the sensitivity of the sensor.
From the result, a flush time at 20\,NL/min of slightly more than \SI{2}{\hour} resulted in a relatively stable $V_c$.
This implied that the uncertainty of the GAr concentration on the HV instability became subdominant. 
From \SI{1.5}{\hour} on until the end of the measurement, the relative change in $V_c$ was 16.4\%.
The absolute breakdown voltage $|V_c|$ plateaued at \SI{18.0}{kV} after \SI{2.26}{\hour} of flushing, corresponding to a bulk field of \SI{5.1}{kV/cm}.
The breakdown voltage between polished spheres in gaseous argon parameterized by the PyBoltz-Townsend model, at a pressure of \SI{1}{atm} and a gap distance of \SI{3.5}{cm}, was reported at \SI{21.7}{kV} \cite{Ar_Xe_breakown:Norman2022}, resulting in a bulk field of \SI{6.2}{kV/cm}.
The obtained lower value from our measurement is expected, as the surface of the plates was not mirror-polished.

\subsubsection{Imaging High-Voltage Breakdown}

An etched mesh identical to the one that was used as the cathode mesh was mounted on the electrode frame, in the same fashion as described in \autoref{sec:hexmesh_assembly}, and used to test the HV setup.
To study potential effects that sharp-edged features could produce under HV, a single hex leg in this mesh was cut and bent downwards towards the ground plate.
This mock-up electrode was then placed and connected to HV inside the test box as describe in \autoref{sec:hv_test_setup}. 
In addition to the two cameras placed above the test box, a third camera was used to videotape the location of the broken leg.

\begin{figure}[htb]
\includegraphics[width=\textwidth]{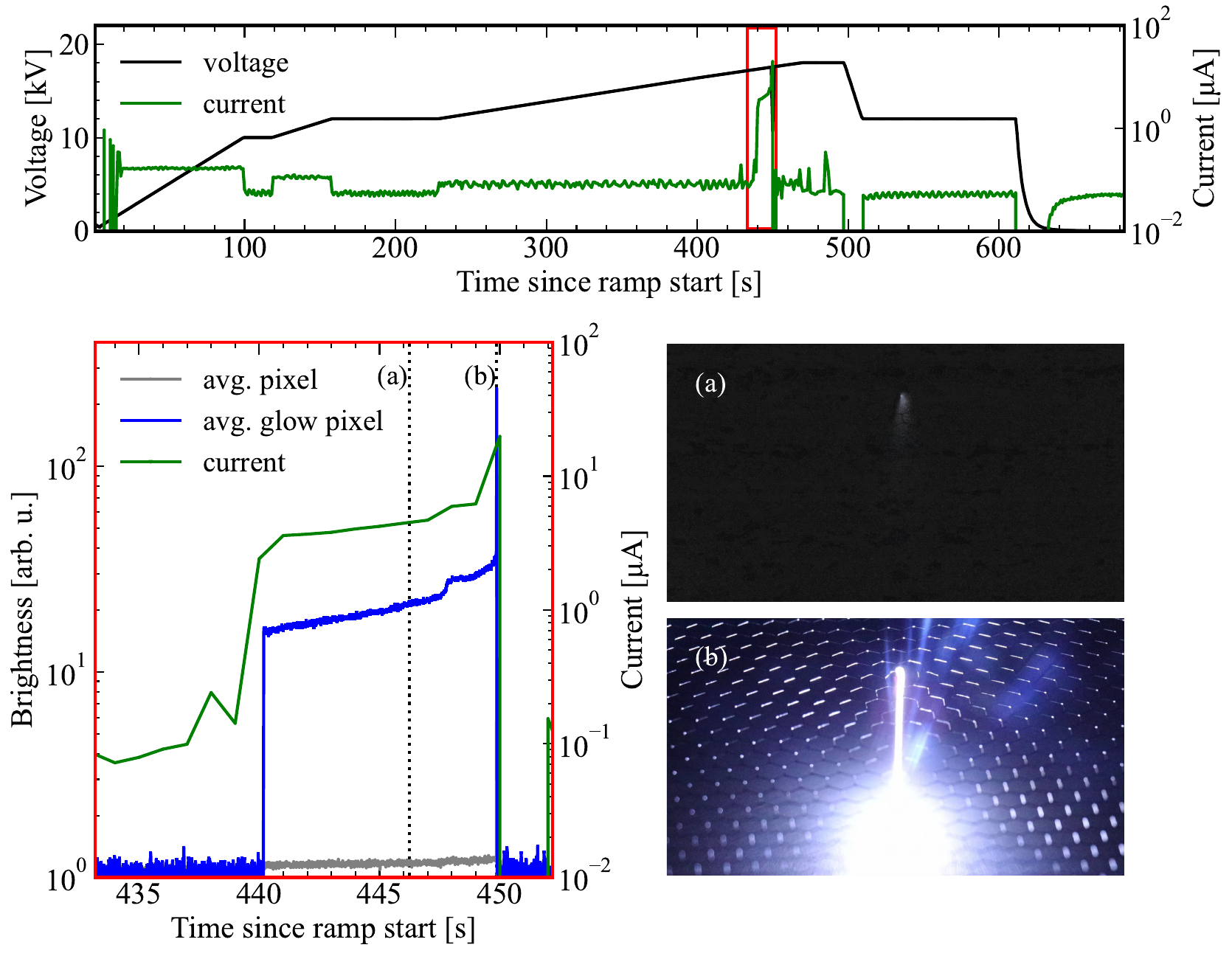}
\caption{Example of the correlation between brightness in the video and the current during an electrical breakdown.
The top panel shows voltage and current as functions of time, with the red square indicating the region that is shown in the lower left plot.
The lower left plot depicts the electrical current and the average pixel value per frame around the time of the electrical breakdown. 
The gray line shows the average pixel value of the full video frame.
The blue line shows the average brightness of a $50\times 25$ pixel region around the position of the broken leg.
It is clearly visible, that these pixels show an increased brightness already before the actual breakdown happens. 
The black dotted lines are placed at the timestamps corresponding to the pictures (a) and (b) on the bottom right.
Picture (a) shows a small glowing spot in the same position where the breakdown happens in picture (b).
Note that for visualization purposes, picture (a) was digitally processed to increase the contrast and thereby make the glow spot visible by eye.
This was not done during the analysis, especially not for the derivation of the blue line in the lower left panel.}
\label{fig:hv_video_correlation}
\end{figure}

As shown in the upper panel of \autoref{fig:hv_video_correlation}, the HV was increased in steps until a breakdown occurred.
The mean brightness of each frame captured by the third camera was used to correlate the light emission to the current profile in \autoref{fig:hv_video_correlation} (lower left).
We computed the average brightness value of all pixels per frame (gray line) and the average value of all pixels in a $50\times 25$ pixel region around the broken leg (blue line).
A gradual increase in the mean brightness was observed at the location of the broken leg, which was directly correlated with the increase in the current profile prior to the breakdown. 
At breakdown the faint glow developed into a continuous arc between the location of the broken hex leg and the ground plate.
The same increase in glow was also observed from the other two cameras that were mounted above the setup with long-exposure images.
This basic test indicated that it was possible to utilize the setup to identify faint glow associated with field emission effects and correlate them with the current profile and specific features on the electrode mesh.

\subsection{High-Voltage Performance}

\begin{figure}[htb]
\includegraphics[width=\textwidth]{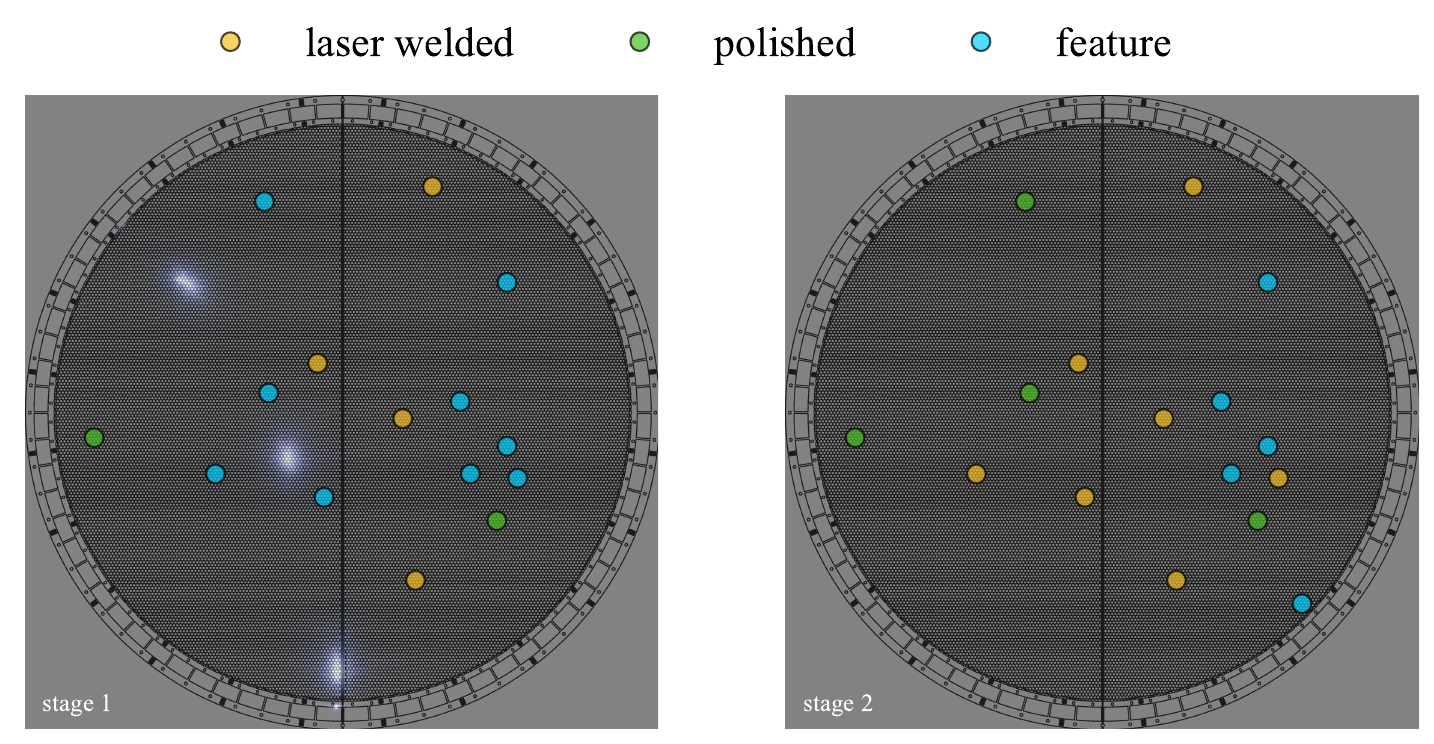}
\caption{Overlay of breakdown images over a design drawing of the electrode mesh including positions of known features.
The yellow and green dots indicate features repaired prior to data taking via laser welding and grinding respectively.
The light blue dots show the positions of known features that were not repaired at the time the data was taken.
The size of the circles is bigger than the uncertainty of the position of the features in the image.
The stage 1 data included 19 breakdown events and only three breakdowns on the mesh itself. 
The rest are not shown as they occurred on the outer part of electrode frame.
At stage 2 testing no breakdowns occurred on the electrode mesh, and only three at the outer part of the electrode frame.
}
\label{fig:mesh_glow_overlay}
\end{figure}

Data that were collected throughout the HV tests listed in \autoref{tab:run_list} were used to assess the performance of the mesh electrode.
Tests that were performed after stage 1 repair resulted in 19 breakdowns, out of which 16 occurred along the stretching ring or on the electrode frame, and only three on the mesh itself which are shown in \autoref{fig:mesh_glow_overlay}.
During HV testing that followed stage 2, only three breakdowns were observed, all on the electrode frame.
The breakdowns at the stretching ring could be potentially attributed to the presence of sharp corners in its geometry as well as lack of electropolishing.
However, as described in \autoref{sec:hexmesh_assembly}, this ring would be removed after the final stretching and installation of the mesh on the electrode frame. 
Breakdowns on the frame could result from sharp edges of the screw heads which were not yet covered as in the final installation. In addition, the aluminum ground plate with its edges not perfectly rounded ended just below the frame.
Breakdowns that occurred on the mesh in stage 1 could have resulted from the presence of dust particles as the tests were performed in a non-cleanroom environment.  
The breakdowns that were observed on the electrode mesh were neither correlated with the locations of known features nor with the locations of features that were repaired via mechanical treatment of laser welding.
Test results described above indicate that both the welding seam and the laser welded repair locations could withstand HV without causing detrimental effects such as field emission or breakdown.

The voltage data were processed and analyzed with the use of custom developed software tools and utilized to evaluate the HV performance of the electrode in GAr.
In each ramp a voltage drop exceeding a threshold of \SI{50}{V} was defined as a breakdown.
All breakdowns without consideration of their spatial distribution on the electrode were taken into account.
Subsequently, the voltage value recorded ten timestamps prior to the breakdown was deemed as the stable voltage and used to calculate the bulk breakdown field assuming a distance of \SI{4.97}{cm} between the electrode mesh and the ground plate.
In cases when multiple consecutive breakdowns were identified within the scope of the same ramp only the first one was used.
In six of the ramps no breakdown were detected, therefore the maximum achieved voltage within the ramp was used to calculate the bulk field.
Such data were deemed as right-censored and were used as an input along the rest of the data in the subsequent statistical analysis.

\begin{figure}[htb]
\centering
\includegraphics[width=.7\textwidth]{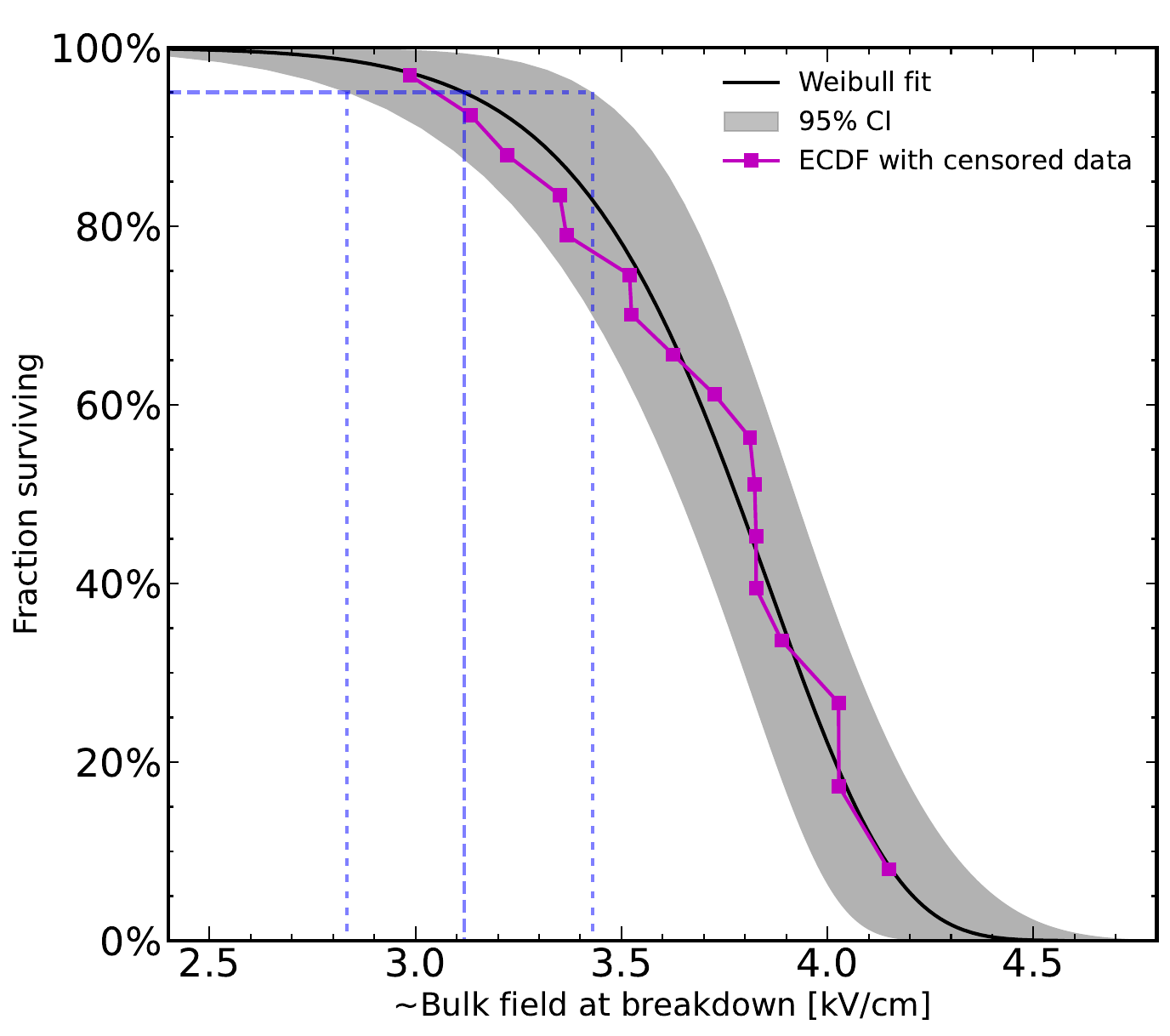}
\caption{Survival probability that was obtained using all suitable data acquired throughout the HV testing of the cathode electrode. Purple squares depict the values of empirical cumulative distribution function (ECDF) of the calculated bulk fields at breakdown, while the survival function that was obtained from the fitted three parameter Weibull distribution is shown in black.
The ECDF is plotted as 1-ECDF, and right-censored data are not shown.
The dashed lines correspond to the lower (left) and upper (right) bounds on the \SI{3.1}{kV/cm} field at 95\% survival probability (middle). 
The gray band corresponds to the 95\% confidence bounds on the bulk field at breakdown obtained with the use of the Reliability \cite{matthew_reid_2020_3938000} fitting framework.}  
\label{fig:survival_plot}
\end{figure}

To quantify the performance of the electrode the obtained breakdown data were fitted with a three parameter Weibull distribution with the use of the Reliability software framework \cite{matthew_reid_2020_3938000}.
The Weibull distribution is suitable for analyzing phenomena whose probability of occurrence below a certain threshold is close to zero \cite{HV_engineering_Faruk}.
Hence, it is commonly employed in reliability analysis in electrical and mechanical engineering to evaluate components and system performance \cite{reliability_modarres}.
The fit results as well as the empirical cumulative distribution function (ECDF) obtained from the data are shown in \autoref{fig:survival_plot}.
From the obtained fit results one can ascertain a bulk field of \SI{3.1}{kV/cm} at 95\% survival probability in GAr, with upper and lower bounds of \SI{2.8}{kV/cm} and \SI{3.4}{kV/cm}, respectively.

Subsequently, we can infer the performance of the electrode in GXe and LXe.
We take the conditions in our setup (GAr, 1.0\,bar, \SI{4.97}{cm} gap distance, and \SI{293}{K}) and typical values in a dual-phase xenon TPC (GXe, 1.8\,bar, \SI{5.4}{cm} gap distance, and \SI{177}{K}). 
Applying Paschen's curves \cite{Ar_Xe_breakown_temp:Massarczyk2016,Ar_Xe_breakown:Norman2022} for the case of uniform fields in both conditions results in a ratio of 1.6 and 2.4 from GAr to GXe.

In a second step, we extrapolate the values from GXe to LXe. 
Although the ionization threshold of LXe was found at around \SI{700}{kV/cm} \cite{LXe_thresholds_Aprile:2014}, breakdowns were recorded at fields lower than this threshold. In a configuration of large electrodes a breakdown field of less than \SI{100}{kV/cm} was reported \cite{lz_electrodes, xebra:2019}.
Various theories have been proposed to account for this difference, such as an earlier breakdown due to the presence of GXe bubbles \cite{LXe_ionization_bubble:1994}.
A field enhancement of a factor~1.2 inside such a bubble cavity is calculated, using the dielectric sphere equation \cite{high_voltage_book:1991} with the dielectric constant of 1.88 for LXe \cite{dielectric_constant_CRC:2017,dielectric_constant:1964}.
Combining this enhancement factor with the conversion factors from GAr to GXe, the conversion from GAr to LXe is at least a factor of 1.9.
As described, the breakdown field at 95\% survival rate is at \SI{3.1}{kV/cm} from our measurement. 
With the minimum conversion factor of 1.9, the hexagonal electrode should withstand a field strength of \SI{5.9}{kV/cm} in LXe.


\section{Conclusions \& Outlook}
\label{sec:conclusions}

In this work we have developed and showcased the capabilities to produce parallel wire and mesh electrodes from simulations and mechanical design to subsequent manufacturing, assembly and HV testing.
For wire electrodes, we have developed a new and improved wire installation technique and successfully utilized it to wire a fully functional  electrode. 
Repairing features in a hexagonal mesh electrode by laser-welding yielded a structurally stable mesh that did not exhibit any observable HV issues. 
Furthermore, a HV test setup was constructed in which we have then tested a hexagonal mesh electrode in GAr atmosphere, reaching field strengths up to 
$\sim$\,\SI{4}{kV/cm}.
Having passed successfully the HV tests in GAr as well as further tests in LXe at the PANCAKE facility \cite{pancake}, the parallel wire grid and the mesh electrode described in this work were installed in the XENONnT TPC as a replacement of the anode and the cathode, respectively.

The next generation of multi-ton xenon-based TPCs such as XLZD \cite{XLZD:2024gxx} will require new approaches when it comes to design, production and assembly of electrodes and other detector components that span several meters in diameter.
The work presented here is part of a comprehensive R\&D program aimed at constructing and testing detector components for the current and next generation of TPCs conducted at Karlsruhe Institute of Technology and the University of Freiburg.
At the University of Freiburg the aforementioned PANCAKE test platform, measuring \SI{2.75}{m} in diameter, was successfully developed and commissioned as a test facility for full scale detector components.
Concurrently, at the Karlsruhe Institute of Technology we pursue a multifaceted R\&D program aimed at developing large-scale electrodes.
Small, centimeter scale samples of electrodes are being studied at the bite-sized High Voltage setup for Electrodes (bHiVE) in GAr, air and vacuum~\cite{Simon2024}.
These studies are aimed at exploring ways for enhancing the HV performance of electrodes via coating, chemical treatment and cleaning techniques, which could be then applied to meter scale electrodes.

Furthermore, fully-automated solutions for both installation and assembly of meter scale electrode grids and their subsequent HV testing are being explored.
A one-meter scale prototype of a robotic system with micrometer precision for HV assessment of electrodes was successfully developed and commissioned. 
This High-voltage Coordinate Unit for Targeted Inspection of Electrodes (HiCUTIE) is currently being upgraded~\cite{Anna2025} to facilitate a fully automated HV assessment of large-scale electrode performance in cleanroom conditions.


\acknowledgments
We thank the technical staff members from local workshops at KIT and University of Freiburg for providing excellent technical solutions in a very cooperative environment.
This work is supported in part by the German Federal Ministry of Research, Technology and Space through grants 05A23VK3, 05A23VF1 and 05A23UM1 in the ErUM-Pro funding line, by the German Research Foundation (DFG) through INST~39/1095-1~FUGG, the Structure and Innovation Fund of the state of Baden-Württemberg (SI-BW), by the Helmholtz Initiative and Networking Fund through grant W2/W3-118 and by the U.S. National Science Foundation grant 2112802. 
A.D. was supported by the Cluster of Excellence “Precision Physics, Fundamental Interactions, and Structure of Matter” (PRISMA$^{+}$ EXC 2118/1) funded by DFG within the German Excellence Strategy (Project ID 390831469).
We appreciate support through the graduate school KSETA at KIT. 
We also gratefully acknowledge support from Laura Baudis (University of Zürich), as well as helpful discussions and exchanges with Jacques Pienaar (Weizmann Institute of Science) and Hans-Werner Ortjohann (University of Münster).

\bibliographystyle{JHEP}
\bibliography{biblio.bib}
\end{document}